\documentclass[a4paper,11pt]{article}
\pdfoutput=1
\usepackage{jcappub} 
\allowdisplaybreaks
\usepackage{float}
\usepackage[dvipsnames]{xcolor}
\usepackage{comment}
\usepackage[normalem]{ulem}

\newcommand{\der}{\mathrm{d}}

\newcommand{\LCDM}{$ \Lambda $CDM~}

\newcommand{\mr}[1]{\textcolor{ForestGreen}{MR: #1}}

\begin{document}

\title{K-mouflage Imprints on Cosmological Observables and Data Constraints}

\author{Giampaolo Benevento$^{a,b}$, Marco Raveri$^{c,d}$,  Andrei Lazanu$^{b,1}$ \note{Now at the D\'epartement de Physique de l'\'Ecole Normale Sup\'erieure, 24 rue Lhomond, 75005 Paris, France}, Nicola Bartolo$^{a,b,e}$, Michele Liguori$^{a,b,e}$, Philippe Brax$^{f}$, Patrick Valageas$^{f}$}

\affiliation{$^{a}$ Dipartimento di Fisica e Astronomia ``G. Galilei'', Universit\`a degli Studi di Padova, via Marzolo 8, I-35131, Padova, Italy}
\affiliation{$^{b}$ INFN, Sezione di Padova, via Marzolo 8, I-35131, Padova, Italy}
\affiliation{$^{c}$ Kavli Institute for Cosmological Physics, Department of Astronomy \& Astrophysics,
Enrico Fermi Institute, The University of Chicago, Chicago, IL 60637, USA}
\affiliation{$^{d}$ Institute Lorentz, Leiden University, PO Box 9506, Leiden 2300 RA, The Netherlands}
\affiliation{$^{e}$ INAF-Osservatorio Astronomico di Padova, Vicolo dell'Osservatorio 5, I-35122 Padova, Italy}
\affiliation{$^{f}$ Institut de Physique Th\'eorique, Universit\`e Paris-Saclay, CEA, CNRS, F-91191 Gif-sur-Yvette, France}

\emailAdd{giampaolo.benevento@phd.unipd.it}


\abstract{
We investigate cosmological constraints on K-mouflage models of modified gravity. 
We consider two scenarios: one where the background evolution is free to deviate from $\Lambda$CDM (K-mouflage) and another one which reproduces a $\Lambda$CDM expansion (K-mimic), implementing both of them into the EFTCAMB code. 
We discuss the main observational signatures of these models and we compare their cosmological predictions to different datasets, including CMB, CMB lensing, SNIa  and different galaxy catalogues. We argue about the possibility of relieving the $H_0$ and weak lensing tensions within these models, finding that K-mouflage scenarios effectively ease the tension on the Hubble Constant. Our final 95\% C.L. bounds on the $\epsilon_{2,0}$ parameter that measures the overall departure from $\Lambda$CDM (corresponding to $\epsilon_{2,0}=0$) are $-0.04\leq \epsilon_{2,0} <0$ for K-mouflage and $0< \epsilon_{2,0} <0.002$ for K-mimic.
In the former case the main constraining power comes from changes in the background expansion history, while in the latter case the model is strongly constrained by measurements of the amplitude of matter perturbations. 
The sensitivity of these cosmological constraints closely matches that of solar system probes.
We show that these constraints could be significantly tightened with future ideal probes like \textit{CORE}.
}


\maketitle
\flushbottom

\section{Introduction}
In the last few decades, a large number of observations have allowed us to test and validate the standard $\Lambda$CDM cosmological model with increasing accuracy. Currently, percent accuracy measurements of $\Lambda$CDM parameters are obtained with \textit{Planck} \cite{2018arXiv180706209P} Cosmic Microwave Background data. Despite its impressive phenomenological success, $\Lambda$CDM presents, however, important open issues. The cosmological constant $\Lambda$ accounts for almost 70\% of the total energy and is a fundamental ingredient to produce the observed late-time cosmic acceleration of the Universe \cite{1998AJ....116.1009R,1999ApJ...517..565P}, but its physical nature remains so far unexplained and its interpretation as vacuum energy is linked to strong naturalness issues. This has prompted theorists to look for  alternative explanations, some of which involve modifications of standard General Relativity (GR), for example via the addition of extra scalar degrees of freedom.

GR has been extensively  and very accurately tested within the Solar System. Therefore, modified gravity (MG) models aiming at explaining cosmic acceleration must in general incorporate a screening mechanism, allowing for standard GR to be recovered on small scales. If we focus our attention on scalar field MG theories that are conformally coupled to matter, there are three known ways to realize screening, respectively via the so called chameleon \cite{2004PhRvL..93q1104K, 2004PhRvD..69d4026K},  Damour-Polyakov \cite{1994NuPhB.423..532D, 2010PhRvD..82f3519B} and derivative screening, the latter being split in Vainshtein \cite{1972PhLB...39..393V} and K-mouflage \cite{2009IJMPD..18.2147B, 2014PhRvD..90b3507B, 2014PhRvD..90b3508B} mechanisms. In this work we focus on 
theories characterized by K-mouflage screening, acting in regions where the gradient of the gravitational potential is higher than a certain threshold. Such theories are built by complementing simple K-essence
scenarios with a universal coupling of the scalar field $\varphi$ to matter \cite{2016JCAP...01..020B}. This 
coupling changes the cosmological dynamics of the models, compared to K-essence, and this generally produces specific signatures already at the background level. These features suggest that K-mouflage theories can be tested very effectively using the CMB, since changes in the expansion history of
the Universe lead to modifications in the angular diameter distance to last scattering, which in turn produce a shift in the position of the peaks of temperature and polarization power spectra. 
Our goal in this paper is therefore that of constraining K-mouflage parameters by using \textit{Planck} data, in a similar fashion to what done for other models, see e.g. \cite{2013PhRvD..88l3514H,2016MNRAS.459.3880H,2017PhRvD..96j3501R}. 
In this paper we also introduce and analyze for the first time a subclass of K-mouflage theories, in which the kinetic term of the scalar degree of freedom is built in such a way as to enforce a quasi-degenerate expansion history with respect to $\Lambda$CDM. We refer to these models as K-mimic scenarios and show that, even in this case, the CMB has strong constraining power, due to changes in the height, rather than the position, of the peaks, and to extra CMB lensing signatures. Other late time probes, such as BAO, are also considered and play a significant role in our analysis. 

The starting step of our analysis is the Effective Field Theory formulation of K-mouflage \cite{2016JCAP...01..020B} and its implementation into the \textsc{EFTCAMB} code \cite{2014PhRvD..89j3530H}. We then employ a Markov-Chain-Monte-Carlo (MCMC) approach \cite{Raveri2014} to place constraints on model parameters, using the \textit{Planck} likelihood \cite{2016A&A...594A...1P}. Interestingly, besides setting stringent constraints, our analysis also shows that K-mouflage and K-mimic can respectively ease the $H_0$ and $\sigma_8$ tension between \textit{Planck} and low redshift probes. 
We also complement our results with Fisher matrix forecasts, showing that the constraints obtained here could be improved in the future by around one order of magnitude with a \textit{CORE}-like CMB survey \cite{2016arXiv161200021D}. Finally, we explicitly show the Horndeski mapping of our theories, which can help in comparing K-mouflage with other MG models and allows to provide direct evidence that gravitational waves travel at the speed of light in K-mouflage.

Our paper is structured as follows. In Section \ref{sec:km} we present the K-mouflage model, its features and its parameterization, also investigating the possibility to reproduce a background evolution degenerate with $\Lambda$CDM. In Section \ref{sec:EFT_theory} we discuss the model in the formalism of the effective field theory of cosmic acceleration and show the mapping of K-mouflage models into Horndeski. In Section \ref{sec:ps}, we use our modified version of EFTCAMB to produce and study the CMB, CMB lensing and matter power spectra in K-mouflage, for different choices of parameters. In Section \ref{sec:constr} we derive MCMC constraints on the parameters of the model and we compute forecasts for future CMB probes.  We draw our conclusions in Section \ref{sec:concl}. Our numerical implementation of K-mouflage in EFTCAMB is further discussed in Appendix \ref{Appendix}.

\section{The K-mouflage model} \label{sec:km}
The K-mouflage class of models with one scalar field, $\varphi$, is defined by the action \cite{2016JCAP...01..020B, 2014PhRvD..90b3507B}
\begin{align}
S &= \int d^4 x \sqrt{-\tilde{g}} \left[ \frac{\tilde{M}_{\rm Pl}^2}{2} \tilde{R} + {\cal M}^4 K(\tilde{\chi}) \right]  +S_{\rm m}(\psi_i, g_{\mu\nu}) \,,
\label{eq:actkm}
\end{align}
where $\tilde{M}_{\rm Pl} = 1/\sqrt{8\pi G}$ is the bare Planck mass, ${\cal M}^4$ is the energy scale of the scalar field, $g_{\mu \nu}$ is the Jordan frame metric, $\tilde{g}_{\mu \nu}$ is the Einstein frame metric,  $g_{\mu \nu}=A^2(\varphi)\tilde{g}_{\mu \nu}$, $\tilde{\chi}$ is defined as
\begin{equation}
\tilde{\chi} = - \frac{\tilde{g}^{\mu\nu} \partial_{\mu}\varphi\partial_{\nu}\varphi}{2 {\cal M}^4}  \,,
\end{equation}
and ${\cal M}^4 K$ is the non-standard kinetic term of the scalar field. 
$S_{m}$ denotes the Lagrangian of the matter fields $\psi_{m}^{(i)}$ that are assumed to be universally coupled to gravity through the Jordan frame metric.
Throughout this paper, we use a `tilde' to denote quantities defined in the Einstein frame.
 
In these theories both the background and perturbation evolution are affected by the universal coupling $A$ and by the scalar field dynamics. 
We parametrize deviations from $\Lambda$CDM at the background level and at linear order in perturbation theory with two functions of the scale factor $a$, related to the coupling $A$ and the kinetic function $K$ by
\begin{equation}
\epsilon_2\equiv \frac{\der \ln \bar{A}}{\der \ln a} \ , \ \ \epsilon_1\equiv \frac{2}{\bar{K}'} \left(\epsilon_2  \tilde{M}_{\rm Pl} \left(\frac{\der \bar{\varphi}}{\der \ln a}\right)^{-1} \right)^2  \,, \label{eps2_func}
\end{equation}
where over bars indicate background quantities and we denote with a prime derivatives with respect to $\tilde{\chi}$, so that $\bar{K}' = \der \bar{K} / \der \tilde{\chi}$.
We follow this notation throughout the paper unless explicitly specified.
As shown in \cite{2016JCAP...01..020B}, the $\epsilon_2$ function governs the running of the Jordan-frame Planck mass $M_{\rm Pl}=\tilde{M}_{\rm Pl}/A$, while $\epsilon_1$ determines the appearance of a late time anisotropic stress and a fifth force. 

Considering linear scalar perturbations around a Friedmann-Lema\^{i}tre-Robertson-Walker (FLRW) background in the Newtonian gauge it can be shown that the Newtonian potential, $\Phi \equiv \delta g_{00}/(2g_{00})$, and intrinsic spatial curvature, $\Psi \equiv -\delta g_{ii}/(2g_{ii})$, on sub-horizon scales and with the quasi-static approximation, are related to gauge-invariant comoving matter density fluctuations through a modified Poisson equation and a modified lensing equation.
These can be written, following \cite{2016JCAP...01..020B}, as
\begin{equation}
\mu \equiv \frac{-k^2 \Phi}{4 \pi G a^2  \bar{\rho}_m \Delta_m} = (1+ \epsilon_1) \bar{A}^2 \, , \ \ \ \Sigma \equiv  \frac{-k^2 (\Phi +\Psi)}{8 \pi G a^2 \bar{\rho}_m \Delta_m} = \bar{A}^2 \ , \label{mu_Sigma}
\end{equation}
where $\bar{\rho}_m$ is the background matter density, the two functions $\mu$ and $\Sigma$ parametrize the departures
from the $\Lambda$CDM evolution of perturbations (given by $\mu=1$ and $\Sigma=1$) at late times. While in general  $\mu$ and $\Sigma$ can be time-  and scale-dependent, for K-mouflage models these two functions only depend on time.

In this paper we normalize the Jordan-frame Planck mass to its current value at $a=1$,
\begin{equation}
A_0 \equiv A(a=1) = 1 \,.
\label{A0-def}
\end{equation}

The action in Eq. (\ref{eq:actkm}) can be used to derive the equations of motion of the scalar field and the Einstein equations, that have been studied in the Einstein frame in \cite{2014PhRvD..90b3507B, 2014PhRvD..90b3508B} and in the Jordan frame in \cite{2015PhRvD..92d3519B}.
Here we recall the background equations of motion in the Jordan frame. The expansion history is described by the K-mouflage Friedmann equations
\begin{equation}
\frac{H^2}{H_0^2} = \frac{\bar{A}^2}{(1-\epsilon_2)^2} \left[ 
\frac{\Omega_{\rm m 0}}{a^3} + \frac{\Omega_{\gamma 0}}{a^4}
+ \Omega_{\varphi 0} \frac{\rho_{\varphi}}{\rho_{\varphi 0}} \right] \ ,
\label{E00-1}
\end{equation}
and
\begin{eqnarray}
- \frac{2}{3 H_0^2} \frac{\der H}{\der t} & = & \frac{\bar{A}^2}{1-\epsilon_2} \;
\left[ \frac{\Omega_{\rm m 0}}{a^3} + \frac{4 \Omega_{\gamma 0}}{3 a^4}
+ \Omega_{\varphi 0} \frac{\rho_{\varphi}+p_{\varphi}}{\rho_{\varphi 0}} \right]
\nonumber \\
&& + \frac{2 \bar{A}^2}{3 (1-\epsilon_2)^2} 
\left( \epsilon_2 - \frac{1}{1-\epsilon_2} \frac{d\epsilon_2}{d\ln a} \right) 
 \left[ \frac{\Omega_{\rm m 0}}{a^3} 
+ \frac{\Omega_{\gamma 0}}{a^4} + \Omega_{\varphi 0}
\frac{\rho_{\varphi}}{\rho_{\varphi 0}} \right] \  .
\label{Eii-1}
\end{eqnarray}

The cosmological density parameters, appearing in the previous equation, are defined for the various species by the ratio between the background energy density $\bar{\rho}_i (a)$ and the critical density $\rho_c (a)= 3 M_{\rm Pl}^2 H^2$, and evaluated  at $z=0$ 
\begin{equation}
\Omega_{\rm m 0} \equiv \frac{\bar{\rho}_{\rm m 0}}{3 \tilde{M}_{\rm Pl}^2 H_0^2} \ , \hspace{0.5cm}
\Omega_{\gamma 0} \equiv \frac{\bar{\rho}_{\gamma 0}}{3 \tilde{M}_{\rm Pl}^2 H_0^2} \ ,  \hspace{0.5cm} \Omega_{\varphi 0} \equiv \frac{\bar{\rho}_{\varphi 0}}{3 \tilde{M}_{\rm Pl}^2 H_0^2} \ ,
\label{continuity}
\end{equation}
respectively for matter, radiation and the scalar field. It turns out that, for a flat spatial curvature, the cosmological density parameters will satisfy: $\Omega_{\rm m } + \Omega_{\gamma}
+ \Omega_{\varphi}= (1-\epsilon_2)^2$, but as shown in \cite{2016JCAP...01..020B}, it is possible to define an effective dark energy density such that $\Omega_{\rm m } + \Omega_{\gamma}
+ \Omega_{DE}= 1$. 
The matter and radiation densities follow the same continuity equation as in $\Lambda$CDM in the Jordan frame, while for the scalar field we have
\begin{equation}
\bar{\rho}_{\varphi} = \frac{{\cal M}^4}{\bar{A}^4} (2 \bar{\tilde\chi} \bar{K}' - \bar{K}) \,, \hspace{0.5cm}
\bar{p}_{\varphi} = \frac{{\cal M}^4}{\bar{A}^4} \bar{K} \,,
\label{rho-phi-def}
\end{equation}
with
\begin{equation}
 \bar{\tilde\chi}  = \frac{\bar{A}^2}{2{\cal M}^4} \left( \frac{\der  \bar{\varphi}}{\der t} \right)^2 \,.
\label{chi_bar}
\end{equation}
To satisfy the Friedmann constraint of Eq.~(\ref{E00-1}) at $z=0$,  using the normalization in Eq.~(\ref{A0-def}), we can write
\begin{equation}
\Omega_{\varphi 0} = (1-\epsilon_{2,0})^2- \left( \Omega_{\rm m 0} + \Omega_{\gamma 0} \right) \,,
\label{Omega-phi0-u0}
\end{equation}
where $\epsilon_{2,0}= \epsilon_2 (a=1)$; this implicitly fixes the value of the scalar field energy scale ${\cal M}^4$. \\
At the background level, the equation of motion of the scalar field 
is equivalent to its continuity equation. In a fashion similar to the $\Lambda$CDM case, we can check that the continuity
equation for the scalar and the two Friedmann equations 
(\ref{E00-1})-(\ref{Eii-1}) are not independent.
Thus, at the background level, one can discard the equation of motion of the 
scalar field and only keep track of the two Friedmann equations. \\
The $\Lambda$CDM limit of the model is recovered when
\begin{equation}
\bar{A}(a)\rightarrow 1, \ \ \epsilon_{2}(a)\rightarrow 0, \ \ \bar{\tilde{\chi}}\rightarrow 0, \ \ \bar{K}'\rightarrow 0  \,, \label{eq:LCDMlimit}
\end{equation}
and the kinetic function in Eq.~(\ref{eq:actkm}) reduces to a cosmological constant. 

\subsection{Reproducing the $\Lambda$CDM expansion history: K-mimic models.} \label{sec:mapping}
The K-mouflage model described in Sec.~\ref{sec:km} usually results in a background expansion history that is different from that of $\Lambda$CDM. 
For the models introduced in Ref.~\cite{2016JCAP...01..020B} the relative deviation in the Hubble function, $H(a)$, is a function of time and model parameters and there is no range for the parameters that allows to produce a $\Lambda$CDM background expansion history without being completely degenerate with the $\Lambda$CDM model at the level of perturbations too. 
This deviation affects different cosmological observables, allowing to constrain the theory already at the background level, as we will show in the next Section. \\

It is worth asking whether it is possible to reproduce a $\Lambda$CDM background evolution, keeping a substantially different dynamics for the perturbations. 
In this Section we explore the possibility to reproduce the same $H(a)$ of $\Lambda$CDM, by appropriately choosing the kinetic function. 
In the following we will refer to this scenario as K-mimic models. \\ 

The K-mouflage model reproduces a $\Lambda$CDM expansion history if the right-hand side of the K-mouflage Friedmann equations (\ref{E00-1})-(\ref{Eii-1})
is equal to the right-hand side of the correspondent $\Lambda$CDM Friedmann equations
\begin{eqnarray}
\frac{H^2}{H_0^2} & = & \frac{\hat{\Omega}_{\rm m 0}}{a^3} + \frac{\hat{\Omega}_{\gamma 0}}{a^4}
+ \hat{\Omega}_{\Lambda 0} \,,
\label{E00-LCDM-1} \\
- \frac{2}{3 H_0^2} \frac{\der H}{\der t} & = & \frac{\hat{\Omega}_{\rm m 0}}{a^3} 
+ \frac{4 \hat{\Omega}_{\gamma 0}}{3 a^4} \,,
\label{Eii-LCDM-1}
\end{eqnarray}
where the cosmological density parameters in the two different models are not assumed to be the same a priori, but we require to recover the same value of $H_0$, so that $\hat{\Omega}_{\rm m 0} \neq \Omega_{\rm m 0}$ implies $\hat{\rho}_{\rm m 0} \neq \rho_{\rm m 0}$ . 
Hence we obtain the equalities
\begin{eqnarray}
\Omega_{\varphi 0} \frac{\rho_{\varphi}}{\rho_{\varphi 0}} & = &
\frac{(1-\epsilon_2)^2}{A^2} \left[ \frac{\hat{\Omega}_{\rm m 0}}{a^3} 
+ \frac{\hat{\Omega}_{\gamma 0}}{a^4} + \hat{\Omega}_{\Lambda 0} \right] - \left( \frac{\Omega_{\rm m 0}}{a^3} + \frac{\Omega_{\gamma 0}}{a^4}
\right) \,,  \label{rho_phi_mimic}
\end{eqnarray}
and
\begin{align}
\Omega_{\varphi 0}  \frac{p_{\varphi}}{\rho_{\varphi 0}} =&\, 
 \frac{1-\epsilon_2}{A^2} \left( -\hat{\Omega}_{\Lambda 0} +
\frac{\hat{\Omega}_{\gamma 0}}{3 a^4} \right) - \frac{\Omega_{\gamma 0}}{3 a^4}
+ \frac{1-\epsilon_2}{3 A^2} \nonumber \\
&\, \times \left( \epsilon_2 + \frac{2}{1-\epsilon_2} 
\frac{\der \epsilon_2}{\der \ln a} \right) \left( \frac{\hat{\Omega}_{\rm m 0}}{a^3} 
+ \frac{\hat{\Omega}_{\gamma 0}}{a^4} + \hat{\Omega}_{\Lambda 0} \right) \, ,  \label{p_phi_mimic}
\end{align}
where for a flat FLRW background ($\hat{\Omega}_{\Lambda 0}= 1-\hat{\Omega}_{\rm m 0} -\hat{\Omega}_{\gamma 0}$), the $\Omega_{\varphi 0}$ parameter is given by Eq.~(\ref{Omega-phi0-u0}).\\
At low redshift, $z \simeq 0$, the scalar field approximately behaves as a cosmological
constant, with $\tilde\chi \ll 1$ and $K \simeq K_0 \simeq -1$.
Therefore, we can choose to normalize the kinetic function such that 
\begin{equation}
\bar{\rho}_{\varphi0}= {\cal M}^4 \Leftrightarrow 2\tilde\chi_0 \bar{K}'_0 - \bar{K}_0 = 1 \,.
\label{N_0}
\end{equation}
Then, using Eqs.~(\ref{rho-phi-def}) we can rewrite Eq.~(\ref{p_phi_mimic}) and Eq.~(\ref{rho_phi_mimic}) as
\begin{align}
\Omega_{\varphi 0} \bar{K} =&\, A^2 (1-\epsilon_2) \left(-\hat{\Omega}_{\Lambda 0} +   
\frac{\hat{\Omega}_{\gamma 0}}{3 a^4} \right) - A^4 \frac{\Omega_{\gamma 0}}{3 a^4}
\nonumber \\
&\, + \frac{A^2(1-\epsilon_2)}{3} \left( \epsilon_2 + \frac{2}{1-\epsilon_2} 
\frac{d\epsilon_2}{d\ln a} \right) \left( \frac{\hat{\Omega}_{\rm m 0}}{a^3} 
+ \frac{\hat{\Omega}_{\gamma 0}}{a^4} + \hat{\Omega}_{\Lambda 0} \right) ,
\label{K_mimic}
\end{align}
and
\begin{eqnarray}
&&\Omega_{\varphi 0} (2 \tilde\chi \bar{K}') =  A^2(1-\epsilon_2) \left(-\hat{\Omega}_{\Lambda 0} +   
\frac{\hat{\Omega}_{\gamma 0}}{3 a^4} \right)
 - A^4 \left( \frac{\Omega_{\rm m 0}}{a^3} 
+ \frac{4 \Omega_{\gamma 0}}{3 a^4}
\right) \nonumber \\ 
&&+ A^2(1-\epsilon_2) \left( 1- \frac{2 \epsilon_2}{3} + \frac{2}{3(1-\epsilon_2)} 
\frac{d\epsilon_2}{d\ln a} \right) \left( \frac{\hat{\Omega}_{\rm m 0}}{a^3} 
+ \frac{\hat{\Omega}_{\gamma 0}}{a^4} + \hat{\Omega}_{\Lambda 0} \right) \ .
\label{K_first_mimic}
\end{eqnarray}
The last two equations can be employed to determine $\bar{\tilde\chi}$ as a function
of the scale factor through
\begin{equation}
\frac{\der\ln\bar{\tilde\chi}}{\der\ln a} =\frac{1}{\bar{\tilde\chi} \bar{K}'} \frac{\der \bar{K}}{\der\ln a} .
\label{chi-a}
\end{equation}
As discussed in Ref.~\cite{2016JCAP...01..020B}, in K-mouflage models one is free to set the present value of both the scalar field $\varphi$ and the kinetic factor $\tilde{\chi}$, corresponding to a choice of normalization for the kinetic function $K(\tilde{\chi})$ and its derivative $K'(\tilde{\chi})$. For K-mimic models, besides the condition given by Eq.~(\ref{N_0}), we impose the normalization $\bar{K}_0'=1$, obtaining the initial
condition for
Eq.~(\ref{chi-a}) at $z=0$, $\tilde\chi_0$, from Eq.~(\ref{K_first_mimic}), while the backward integration provides $\tilde\chi(a)$ at all times.
Together with Eq.~(\ref{K_mimic}), this gives a parametric definition of the kinetic
function $K(\tilde\chi)$. \\
To complete the definition of the K-mimic model, we implicitly define the
conformal coupling through a given function $\bar{A}(a)$. This directly yields
the factor $\epsilon_2(a)$ from Eq.~(\ref{eps2_func}), and we obtain $\varphi(a)$
by integrating Eq.~(\ref{chi_bar}), with the initial condition
$\varphi(t=0)=0$. 
This provides a parametric definition of the coupling $A(\varphi)$. \\
To obtain a background evolution completely degenerate with a $\Lambda$CDM model, we should impose $\hat{\Omega}_i= \Omega_i$ for all species. However, this requirement does not satisfy the stability conditions discussed in \cite{2016JCAP...01..020B} to avoid ghosts.
Indeed, as we require $\tilde\chi > 0$, $K'>0$ and $A > 0$ we can see from 
 Eq.~(\ref{K_first_mimic}) and Eq.~(\ref{eps2_func}) that we must have
\begin{equation}
\frac{\bar{A}}{3a^4} \left[(3a \hat{\Omega}_{m0}+ 4\hat{\Omega}_{\gamma 0})\left( \bar{A}-a\frac{\der \bar{A}}{\der a} \right)  - (3a \Omega_{m0}+ 4\Omega_{\gamma 0})\bar{A}^3+ 2a^2(a^4 \hat{\Omega}_{\Lambda} +a \hat{\Omega}_{m0} + \hat{\Omega}_{\gamma 0}) \frac{\der^2 \bar{A}}{\der a^2} \right] > 0 \ \ .
\label{K-mimic-constraint}
\end{equation}
This inequality must be satisfied in the range $0 \leq a \leq 1$. Indeed, using the normalization Eq.~(\ref{A0-def}) for the coupling function, and taking $\epsilon_2 >0$, the left hand side of Eq.~(\ref{K-mimic-constraint}) is a decreasing function of $a$. Imposing 
$\hat{\Omega}_{\gamma 0}= \Omega _{\gamma 0}$, as both the parameters are fixed by measurement of the CMB temperature, we are left with a condition on the parameter $\hat{\Omega}_{\rm m0}$ at $a=1$
\begin{equation}
\hat{\Omega}_{\rm m0} > \frac{\Omega_{\rm m0}}{1-\epsilon_{2,0}} + 4 \Omega_{\gamma 0} \frac{ \epsilon_{2,0} - 2 \frac{\der^2 \bar{A}}{\der a^2}\vert_{a=1}}{3(1-\epsilon_{2,0})}  \ . 
\label{Omega-constraint}
\end{equation}
Equation~(\ref{Omega-constraint}) shows that even within K-mimic models, the background evolution cannot be completely degenerate with $\Lambda$CDM. Indeed, given a set of cosmological parameters \{$ \Omega_{\rm b0}, \ \Omega_{\rm c0}, \ \Omega_{\gamma0}, \ H_0$\} K-mimic models reproduce the same $H(a)$ of a $\Lambda$CDM model with a slightly higher matter density. \\ Once a value for $\hat{\Omega}_{\rm m0}$ is picked, in agreement with the condition in Eq.~(\ref{Omega-constraint}), this automatically fixes the present value of $\tilde\chi$ via  Eq.~(\ref{K_first_mimic}). At $z=0$ we should have $\tilde\chi \ll 0$ to recover a cosmological constant behaviour, so a natural choice is to take $\tilde\chi_0 \sim \epsilon_{2,0}$, allowing to recover the exact $\Lambda$CDM behaviour if $\epsilon_{2,0} \rightarrow 0$. Our specific choice for $\tilde\chi_0 $ and  $\hat{\Omega}_{\rm m0}$ is reported in Eq.~(\ref{Omega-m-value}) of Appendix \ref{Appendix}.

\subsection{Parametrization of the models}\label{sec:parameters}
In order to test K-mouflage against cosmological observations, we define the coupling function and the kinetic term as functions of the scale factor in terms of a set of parameters which will be varied together with the standard cosmological parameters. The solution  of the background evolution equations for the model provides the relation between $\tilde{\chi}$, $\varphi$ and $a$, allowing to reconstruct the $K(\tilde{\chi})$ and $A(\varphi)$ functions defined in the action (\ref{eq:actkm}). \\ We consider two different scenarios: a five-function parametrization of K-mouflage introduced in \cite{2016JCAP...01..020B} and a three parameter formulation of K-mimic models defined in Sec.~\ref{sec:mapping}. In both cases the background coupling functions is defined in terms of three parameters $\epsilon_{2,0}$, $\gamma_A$, $m$ as 
\begin{equation}
 \bar{A}(a)=1+\alpha_A- \alpha_A \left[ \frac{a (\gamma_A+1)}{a+\gamma_A} \right]^{\nu_A} \label{A_def}  \, ,
\end{equation}
with 
\begin{align}
\nu_A&=\frac{3 (m-1)}{2m-1} \, , \\
\alpha_A&=-\frac{\epsilon_{2,0}(\gamma_A+1)}{\gamma_A \nu_A} \, .
\end{align}
The kinetic function for K-mimic models is given by Eq.~(\ref{K_mimic}), requiring no additional parameters. 
For K-mouflage models the kinetic function can be computed integrating the following expression for its derivative
\begin{align}
\frac{\der \bar{K}}{\der \tilde{\chi}}&= \frac{U(a)}{a^3 \sqrt{\bar{\tilde{\chi}}}} \,\label{eq:dK/dchi} , \\
U(a) &= U_0 \left( (\sqrt{a_{\mathrm{eq}}}+1)+ \frac{\alpha_U}{\ln(\gamma_U+1)} \right) \frac{a^2 \ln(\gamma_U+a)}{(\sqrt{a_{\mathrm{eq}}}+\sqrt{a}) \ln(\gamma_U+a)+ \alpha_U a^2} \label{U_def}  \, , \\
\sqrt{\bar{\tilde{\chi}}}& =-\frac{\bar{\rho}_{m0}}{{\cal M}^4} \frac{\epsilon_2 \bar{A}^4}{2 U(-3 \epsilon_2+\frac{d \ln U}{d \ln a})} \, \label{chi_tilde_U} ,
\end{align}
where we have introduced two additional parameters $\alpha_U$ and $\gamma_U$, while  $a_{\mathrm{eq}}$ represents the scale factor at radiation-matter equality  and the normalization $U_0$ is given in Appendix \ref{Appendix}.\\
The allowed range of the parameters is restricted to fit the natural domain of the two functions $U(a)$ and $\bar{A}(a)$ and additional constraints that ensure the stability of the solutions have to be satisfied.  Specifically, as discussed in Ref. \cite{2016JCAP...01..020B}, all K-mouflage models must satisfy the conditions
\begin{equation}
\bar{K}'>0, \ \ \bar{A}>0, \ \ \bar{K}'+ 2\tilde{\chi}\bar{K}''>0,
\label{stability_conditions}
\end{equation}
as well as the Solar System and cosmological constraints \cite{2015PhRvD..91l3522B}. For a more clear interpretation of the results, let us recall the physical meaning of the different parameters and the bounds they have to satisfy.
\begin{itemize}
\item  $\epsilon_{2,0}$; this parameter sets the value of the $\epsilon_2$ function today. The $\Lambda$CDM limit is recovered when $\epsilon_{2,0} \to 0$, independently of the values of the other four parameters. For K-mouflage models, adopting the same convention as \citep{2016JCAP...01..020B} we choose this parameter to be negative. Conversely in the case of K-mimic models $\epsilon_{2,0}$ has to be positive in order to match the stability requirement. As shown in \cite{2015PhRvD..91l3522B}, Solar System tests impose $|\epsilon_{2,0}|\lesssim 0.01$. In our analysis we do not use an informative prior on this parameter, as we want to compare cosmological constraints with Solar System bounds. 
\item $m>1$; describes the large $\tilde{\chi}$ behaviour of the kinetic function. It is possible to show \citep{2016JCAP...01..020B} that, given the parametrization described by Eqs.~(\ref{A_def})-(\ref{chi_tilde_U}), in the limit of large $\tilde{\chi}$ the kinetic term follows the asymptotic power-law behaviour: $K(\tilde{\chi}) \sim \tilde{\chi}^{m}$. 
As done in previous works, in some plot of Sec.~\ref{sec:ps} we study the particular case called ``cubic model'' which is obtained by fixing $m= 3$.
\item $\gamma_A>0$; describes the transition to the dark energy dominated epoch in the $A(a)$ coupling function. Natural values for this parameter are of order unity \cite{2016JCAP...01..020B}. As discussed in Sec.~\ref{sec:ps} we verified that high values for this parameter push the model toward the $\Lambda$CDM limit, however values of $\gamma_A \gtrsim 20$ are likely to be excluded by the stability conditions in Eq.~(\ref{stability_conditions}).
\item $\gamma_U \ge 1$ and $\alpha_U>0$; these two parameters set the transition to the dark energy dominated epoch in the $K(a)$ kinetic function. We checked that early time probes (like CMB temperature anisotropies), as well as late time probes (CMB lensing) are practically insensitive to the parameter $\gamma_U$ which can be safely fixed to 1, i.e. the minimum value that that avoids negative values of the $U(a)$ function. The parameter $\alpha_U$ has some influence on late-time probes on large scales, as we will show in Sec.~\ref{sec:ps}.
\end{itemize}
Although there are no a priori upper bounds on the parameters, by investigating the numerical behaviour of the solutions to the equations, we have checked that if too high values of these parameters are taken, either there are negligible changes in the results or ghosts appear. Summarizing, we have taken the parameters to be in the range specified by Table~\ref{table:parameters}.
\begin{table}
\begin{center}
\begin{tabular}{l | ccc}
\multicolumn{3}{c}{Allowed range of the parameters}\\
\hline
Parameter & K-mouflage & K-mimic \\
\hline
$\epsilon_{2,0}$ & $[-1, \ 0.0]$ & $[0.0, \ 1.0]$ \\
$\gamma_A$ &  $[0.2, \ 25]$ &  $[0.2, \ 25]$ \\
$m$ & $ [1, \ 10] $ & $[1, \ 10] $ \\
$\gamma_U$ &  $[1, \ 10] $ & $\diagdown$ \\
$\alpha_U$  & $[0, \ 2] $ & $\diagdown$ \\
\end{tabular}
\caption{
\label{table:parameters}
The range for the K-mouflage and K-mimic parameters assumed in our analysis. 
}
\end{center}
\end{table}
\section{K-mouflage in the Effective Field Theory of Dark Energy}
\label{sec:EFT_theory}
The EFT of dark energy represents a general framework for describing dark energy and modified gravity that includes all single field models \cite{2013JCAP...02..032G, 2013JCAP...08..010B, 2013JCAP...08..025G, 2013JCAP...12..044B, 2014JCAP...05..043P}. It is built in the unitary gauge in analogy to the EFT of inflation \cite{2006JHEP...12..080C, 2008JHEP...03..014C} by using operators represented by perturbations of quantities which are invariant under time dependent spatial diffeomorphisms: $g^{00}$, the extrinsic curvature tensor $K^{\mu}_{\; \nu}$ and the Riemann tensor $R_{\mu \nu \rho \sigma}$. \\
The mapping of K-mouflage into the EFT formalism  has been presented in \cite{2016JCAP...01..020B} and here we will briefly summarize the main steps and the final result.\\
Starting from the action given in Eq. (\ref{eq:actkm}), we can make a change of coordinates to the unitary gauge, for which constant time hypersurfaces coincides with the uniform $\varphi$-hypersurfaces. Then, by definition, the scalar field only depends on time and so do the coupling function and the kinetic function. Hence, one can easily write the action in the Jordan frame, and expand it in perturbations of the time-time component of the metric tensor, around its value on a FLRW background. Now the action is directly comparable with the EFT one  \cite{2013JCAP...02..032G} and can be expressed as

\begin{align}
S = & \int d^4 x \sqrt{-g} \left[ \frac{M_{\rm Pl}^2}{2} R
- \Lambda(\tau) - c(\tau) g^{00} \right. 
 \left. + \sum_{n=2}^{\infty}
\frac{M_n^4(\tau)}{n!} (\delta g^{00})^n \right] +S_{\rm m}(\psi_i, g_{\mu\nu}) \,,
\end{align}
with
\begin{align}
M_{\rm Pl}^2 =&\, \bar{A}^{-2} \tilde{M}_{\rm Pl}^2  \nonumber \\
\Lambda =&\, - \bar{A}^{-4} {\cal M}^4 (\bar{K}+ \chi_{*} \bar{K}' \tilde{g}^{00}) \nonumber \\
c =&\, A^{-2} \tilde{c} -\frac{3}{4} M_{\rm Pl}^2 \left(\frac{d \ln (\bar{A}^{-2})}{d \tau} \right)^2 \nonumber \\
M_n^4 =&\, \bar{A}^{-2(2-n)} {\cal M}^4 (-\chi_{*})^n \bar{K}^{(n)} \,,
\end{align}
where $\bar{K}^{(n)} \equiv \der^n \bar{K}/\der \tilde{\chi}^n$, $\chi_{*}=\bar{A}^{-2}\chi$ and the overbar denotes background quantities. We see that only the three EFT functions that regulate the background evolution, together with operators involving perturbations of $g^{00}$, appear.
This mapping allows the K-mouflage model to be incorporated into the EFTCAMB code \cite{2014PhRvD..89j3530H,Raveri2014} in order to compute the cosmological observables of interest.

%
\subsection{K-mouflage models and Horndeski models}
\label{sec:interpret}

The EFT approach, discussed in the previous Section, is a powerful and universal way of describing dark energy and modified gravity models. \\
In this subsection we consider another class of modified gravity models, namely Horndeski \cite{1974IJTP...10..363H}, which encompasses all single-field models with at most second order derivatives in the resulting equation of motion.
In Ref. \cite{2013JCAP...08..025G} it has been shown that in the case of the Horndeski class of actions, besides one function for the background, only four functions of time are required to describe fully linear perturbation theory.

Using the parametrization introduced in \cite{Bellini2014}, these functions are labelled as: $\alpha_K$ -- kineticity, $\alpha_B$ -- braiding, $\alpha_M$ -- running of the Planck mass, $\alpha_T$ -- tensor excess speed.

We aim to discuss the properties of the perturbations of the K-mouflage models in this general framework, by expressing Eq. \ref{eq:actkm} in the Jordan frame and matching the terms to the general form
\begin{equation}
 S = \int d^4 x \sqrt{-g} \left[\sum_{i=2}^5{\cal L}_{i}+{\cal L}_{m}[g_{\mu \nu}] \right] \,,
\end{equation}
with
\begin{align}
{\cal L}_{2} & =  K_H (\varphi,\, X) \nonumber\\
{\cal L}_{3} & =  -G_{3}(\varphi,\, X)\Box\varphi \nonumber\\
{\cal L}_{4} & =  G_{4}(\varphi,\, X)R+G_{4X}(\varphi,\, X)\left[\left(\Box\varphi\right)^{2}-\varphi_{;\mu\nu}\varphi^{;\mu\nu}\right] \nonumber\\
{\cal L}_{5} & =  G_{5}(\varphi,\, X)G_{\mu\nu}\varphi^{;\mu\nu}-\frac{1}{6}G_{5X}(\varphi,\, X)
\left[\left(\Box\varphi\right)^{3}+2{\phi_{;\mu}}^{\nu}{\varphi_{;\nu}}^{\alpha}{\varphi_{;\alpha}}^{\mu}-3\phi_{;\mu\nu}\varphi^{;\mu\nu}\Box\varphi\right] \,. \label{eq:Hordenski}
\end{align}

Hence, in the K-mouflage theories, the terms appearing in the action (Eq. \ref{eq:actkm}) of the Horndeski action are given by
\begin{align}
 K_H &= \frac{{\cal M}^4}{A^4} K \left( \tilde{\chi} \right) +6 \tilde{\chi} \frac{{\cal M}^4 }{A^5} \tilde{M}_{\rm Pl}^2 \left(2 \left( \frac{\der A}{\der \varphi}\right)^2 \frac{1}{A}-\frac{\der^2 A}{\der \varphi^2} \right) \nonumber\\
 G_3 &=-3 \tilde{M}_{\rm Pl}^2 \left( \frac{\der A}{\der \varphi}\right) \frac{1}{A^3} \nonumber\\
 G_4 &=\frac{1}{2} \frac{\tilde{M}_{\rm Pl}^2}{A^2} \nonumber\\
 G_5 &=0  \,,
\end{align}
where $K_H$ is the Horndeski function defined in the first line of Eq.~(\ref{eq:Hordenski}). The variable $X$ satisfies $\tilde{\chi}=X A^2 /{\cal M}^4 $. Based on this mapping, we derive the coefficients $\alpha_i$ of Ref. \cite{Bellini2014} for the general K-mouflage model,
\begin{align} \label{eq:alphaT}
 \alpha_K &= \frac{2{\cal M}^4}{A^2} \frac{\tilde{\chi}}{H^2} \left[-6  \left( \frac{\der A}{\der \varphi}\right)^2 \frac{1}{A^2}+\frac{K'( \tilde{\chi})}{\tilde{M}_{\rm Pl}^2}+\frac{2 K''( \tilde{\chi})}{\tilde{M}_{\rm Pl}^2}\tilde{\chi} \right] \nonumber \\
 \alpha_B &= \frac{2}{H} \frac{1}{A} \left( \frac{\der A}{\der \varphi}\right) \dot{\varphi}   \nonumber \\
 \alpha_M &= -\frac{2}{H} \frac{1}{A} \left( \frac{\der A}{\der \varphi}\right) \dot{\varphi}   = - \alpha_B  \nonumber \\
 \alpha_T &= 0  \, ,
\end{align}
where primes denote derivative with respect to $\tilde{\chi}$ and the overdot denotes derivative with respect to proper time. 
Using the solutions of Sec.~\ref{sec:km}, these general functions can be expressed in terms of the explicit parametrisation of \cite{2016JCAP...01..020B}
\begin{equation}
 \alpha_B = 2 \epsilon_2 = -  \alpha_M \, , \label{eq:alphaB2}
\end{equation}
while $\alpha_K$ can be calculated using Eqs.~(\ref{eps2_func}), (\ref{chi_bar}), and (\ref{eq:dK/dchi}) in terms of $\epsilon_2$, $U$, and their derivatives with respect to the scale factor, thus all the $\alpha$-functions can be also written in terms of the parameters introduced in Sec~\ref{sec:parameters}.

Eq.~(\ref{eq:alphaT}) shows that gravitational waves travel at the speed of light, while Eq. (\ref{eq:alphaB2}) shows that for K-mouflage models with $\epsilon_{2,0}<0$ braiding is small and negative, $|\alpha_B| \lesssim {\cal O}\left(10^{-2}\right)$, while the running of the effective Planck mass is small and positive, the opposite holds for K-mimic models with $\epsilon_{2,0}>0$. 
The kineticity is not expected to modify significantly the growth of matter or of metric perturbations on sub-horizon scales with respect to standard GR~\cite{2013JCAP...08..010B,Pogosian:2016pwr,2016JCAP...07..040R}, but it can affect super-horizon scales, and generate an observable effect when those scales enters the horizon at late times. \\ 
The running of the effective Planck mass and the braiding are both known to affect the evolution of the Bardeen potentials and the matter fluctuations in a non-trivial and scale-dependent way. The non-zero $\alpha_M$ also generates a late-time anisotropic stress, in agreement with the results of Sec.~\ref{sec:km}. The combination of these effects determines changes in the matter and lensing power spectra, as we will show in the next Section. The braiding and the running of the Planck mass are also expected to influence the Integrated Sachs-Wolfe (ISW) effect. As we discuss in the next-section, we expect a significant enhancement of the early-ISW for K-mimic models.  
Even though it is sub-dominant in the CMB temperature-temperature anisotropy spectrum,  this signature can be explored through cross-correlation between CMB temperature and galaxy number counts, which constitutes an important test for these models \cite{Benevento2018}.

\section{Power Spectra}
\label{sec:ps}

\begin{figure}[tbp]
\begin{center}
\includegraphics[width=\linewidth]{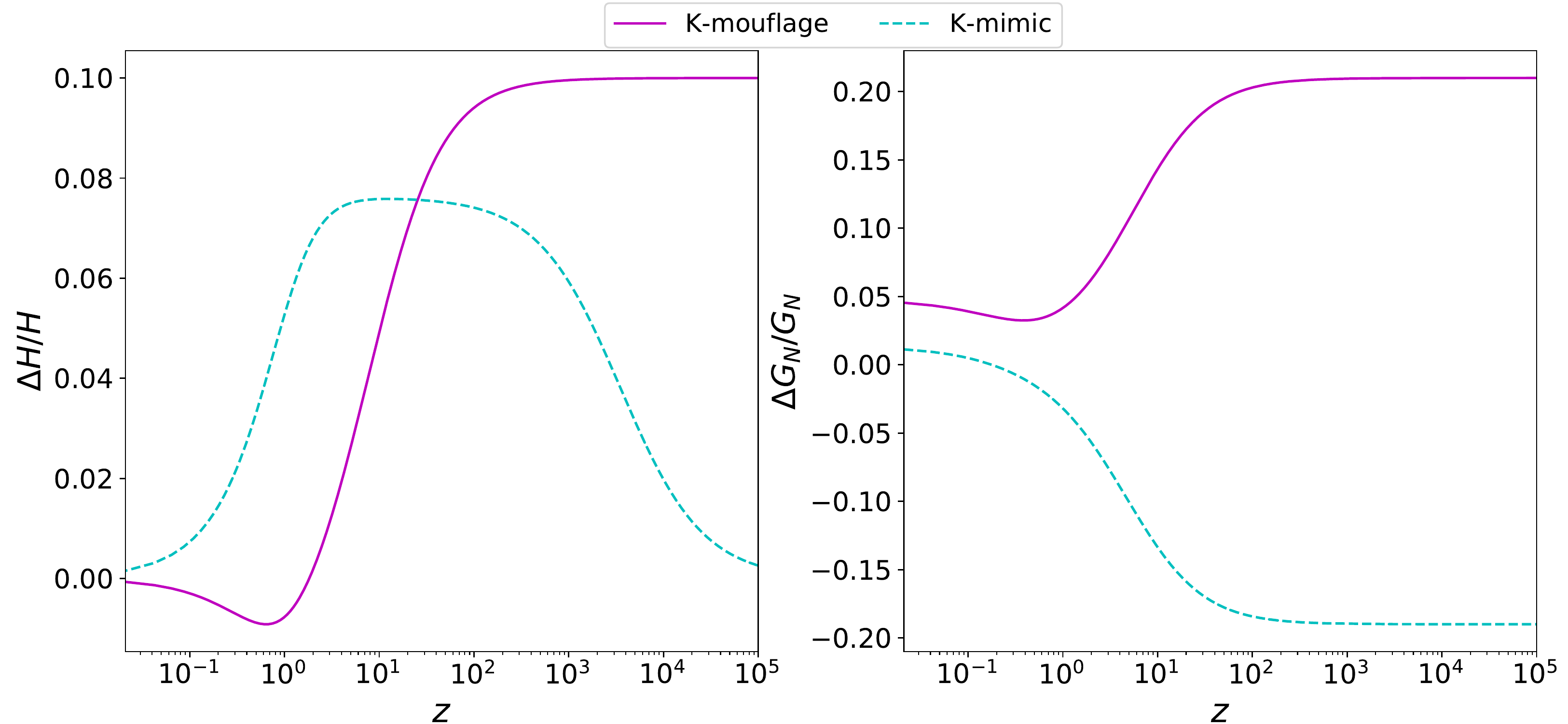}
\caption{ \emph{Left panel}: Relative deviation of the Hubble function $\Delta H/ H$ from the $\Lambda$CDM reference. \emph{Right panel}: Relative deviation of the effective Newtonian constant, from the $\Lambda$CDM reference. The effective Newtonian constant is defined as $G_{N,eff}= \mu G_N$, with $\mu$ given in Eq.~[\ref{mu_Sigma}]. We consider a K-mouflage model with parameters \{$\alpha_U=1$, $\gamma_U=1$, $m=3$, $\gamma_A=0.2$, $\epsilon_{2,0}=-2\times 10^{-2}$\} and a K-mimic model with parameters  \{$m=3$, $\gamma_A=0.2$, $\epsilon_{2,0}=2\times 10^{-2}$\}.   As we can see, K-mimic models reproduce the expansion history given by $H^2 = H_0^2 (\hat{\Omega}_{m,0}/a^3+ \Omega_{\gamma,0}/a^4 +( 1-\hat{\Omega}_{m,0}- \Omega_{\gamma,0}))$ and recover the $\Lambda$CDM solution in this plot, given by  $H^2 = H_0^2 (\Omega_{m,0}/a^3+ \Omega_{\gamma,0}/a^4 + ( 1-\Omega_{m,0}- \Omega_{\gamma,0}))$, for $a \ll a_{eq}$. K-mouflage shows instead substantial deviations in the background expansion, throughout all the cosmic epochs.
}
\label{fig:adotoa}
\end{center}
\end{figure}
\begin{figure}[tbp]
\begin{center}
\includegraphics[width=\linewidth]{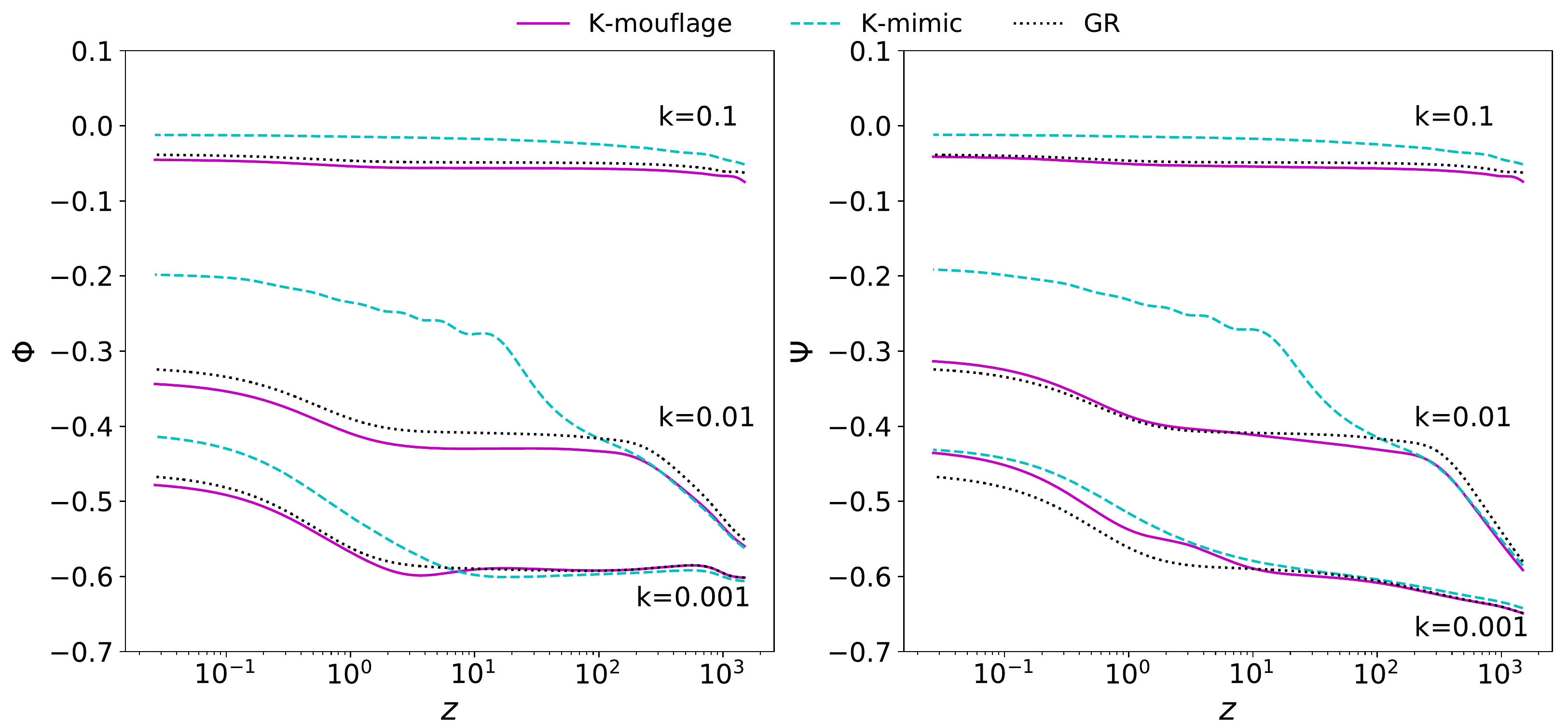}
\caption{Evolution of the Bardeen potentials $\Phi$ and $\Psi$, as defined in Sec~\ref{sec:km}, for three different Fourier modes in K-mouflage, K-mimic and GR. The mode $k=0.1$ enters the horizon at $z \sim 4 \times 10^4$, the mode $k=0.01$ enters the horizon at $z \sim 3200$, while the mode $k=0.001$ enters the horizon at $z \sim 5.6$. We use the same parameters of the previous plot for K-mouflage and K-mimic.}
\label{fig:Psi_Phi}
\end{center}
\end{figure}
\begin{figure}[tbp]
\begin{center}
\includegraphics[width=\columnwidth]{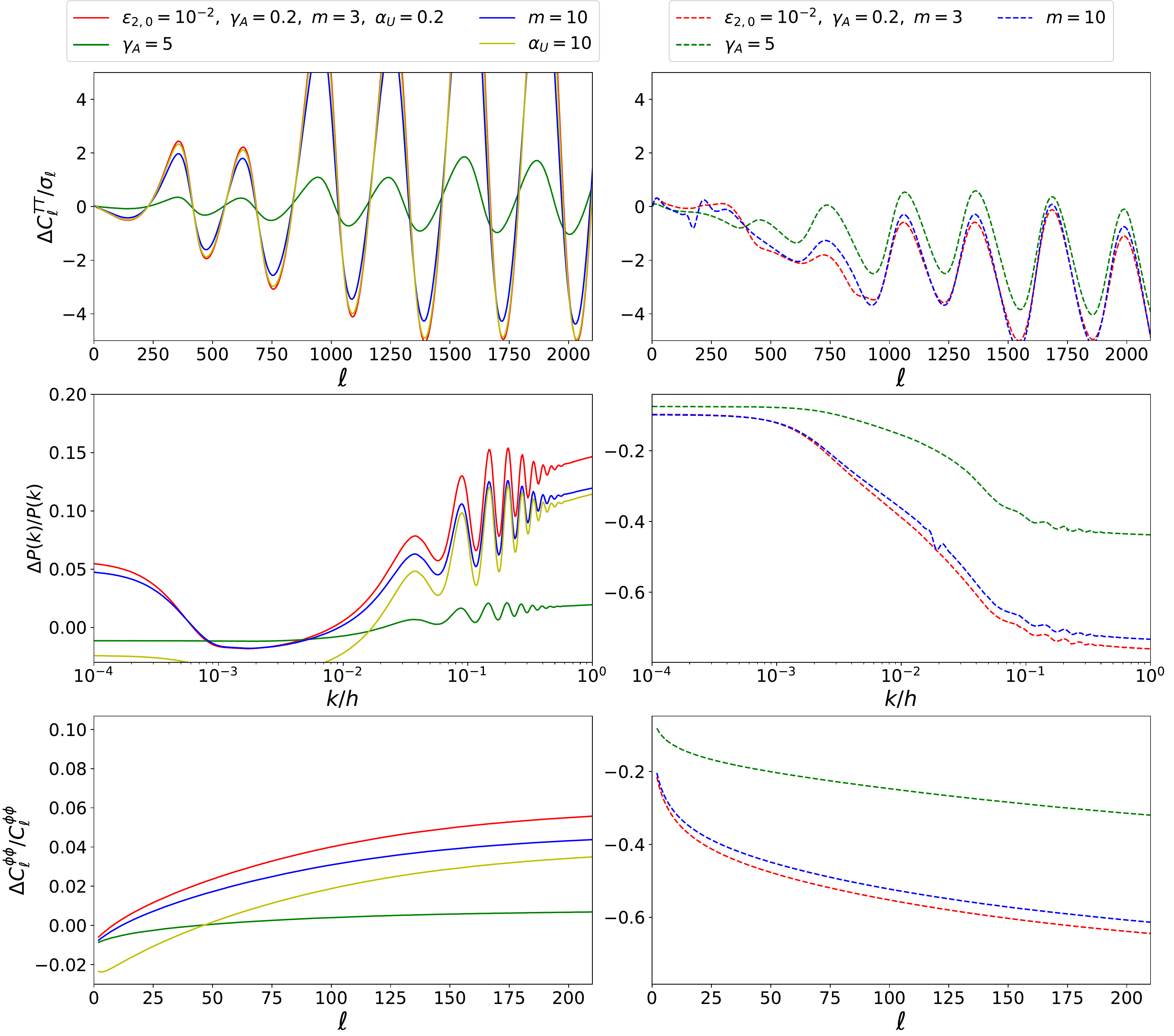}
\caption{Effect of different K-mouflage and K-mimic parameters on cosmological observables. \emph{Upper panels}: relative deviation of the CMB temperature anisotropies power spectrum from the $\Lambda$CDM prediction in units of its variance per multipole
 $\sigma_\ell = \sqrt{2/(2 + \ell)}C_{\ell}^{TT(\Lambda CDM)}$. \emph{Middle panels}: relative deviation of the matter power spectrum from the $\Lambda$CDM prediction $\Delta P(k)/P(k)^{\Lambda CDM}$. \emph{Lower panels}: relative deviation of the CMB lensing potential power spectrum from the $\Lambda$CDM prediction  $\Delta C_{\ell}^{\Phi \Phi} /C_{\ell}^{\Phi \Phi(\Lambda CDM)}$. We show K-mouflage models (left panels, continuous lines) and K-mimic models (right panels, dashed lines) with different choice of the parameters in agreement with Solar System constraints (i.e. they have $\vert \epsilon_{2,0} \vert=0.01$). Taking the red line as reference, we change one parameter per time, obtaining the models labelled with different colours. The parameter $\gamma_U$ is fixed to $1$ for all the K-mouflage models.  }
\label{fig:rel_dev}
\end{center}
\end{figure}

We have used our version of EFTCAMB \cite{2014PhRvD..89j3530H} to compute the CMB power spectrum in the full K-mouflage and in the K-mimic models, for different values of the parameters. In this Section we discuss the effect of varying parameters on the cosmological observables. For all the models shown in the plots, we fix the baryon density
at $\Omega_b h^2=0.0223$, the dark matter density at $\Omega_{c} h^2=0.119$,
the reduced Hubble constant at $h = 0.67$, the spectral index at 
$n_s = 0.965$, the initial amplitude of comoving curvature fluctuation
at $A_s = 2.1 \times 10^{-9}$ ($k_0 = 0.05 \rm Mpc^{-1}$) and the reionization optical depth at $\tau = 0.05$. \\ The combined effect of the running of the Planck mass and of the fifth force, alters gravity at early and late times. This affects both the cosmological background and the perturbation dynamics. \\ For K-mouflage models the expansion history deviates from $\Lambda$CDM, also at early times during the radiation dominated epoch. K-mimic models produce the expansion history of a $\Lambda$CDM model with an increased matter density ($\hat{\Omega}_m$), as explained in Sec.~\ref{sec:mapping}. This implies that, for a fixed matter density, the Hubble rate deviates during the matter-dominated epoch while it recovers the $\Lambda$CDM solution during the radiation-dominated era. This behaviour is displayed in the left panel of Fig.~\ref{fig:adotoa}, where we plot the relative deviation of the Hubble of rate from the $\Lambda$CDM reference for two representative K-mouflage and K-mimic models.
\begin{figure}[tbp]
\begin{center}
\includegraphics[width=\linewidth]{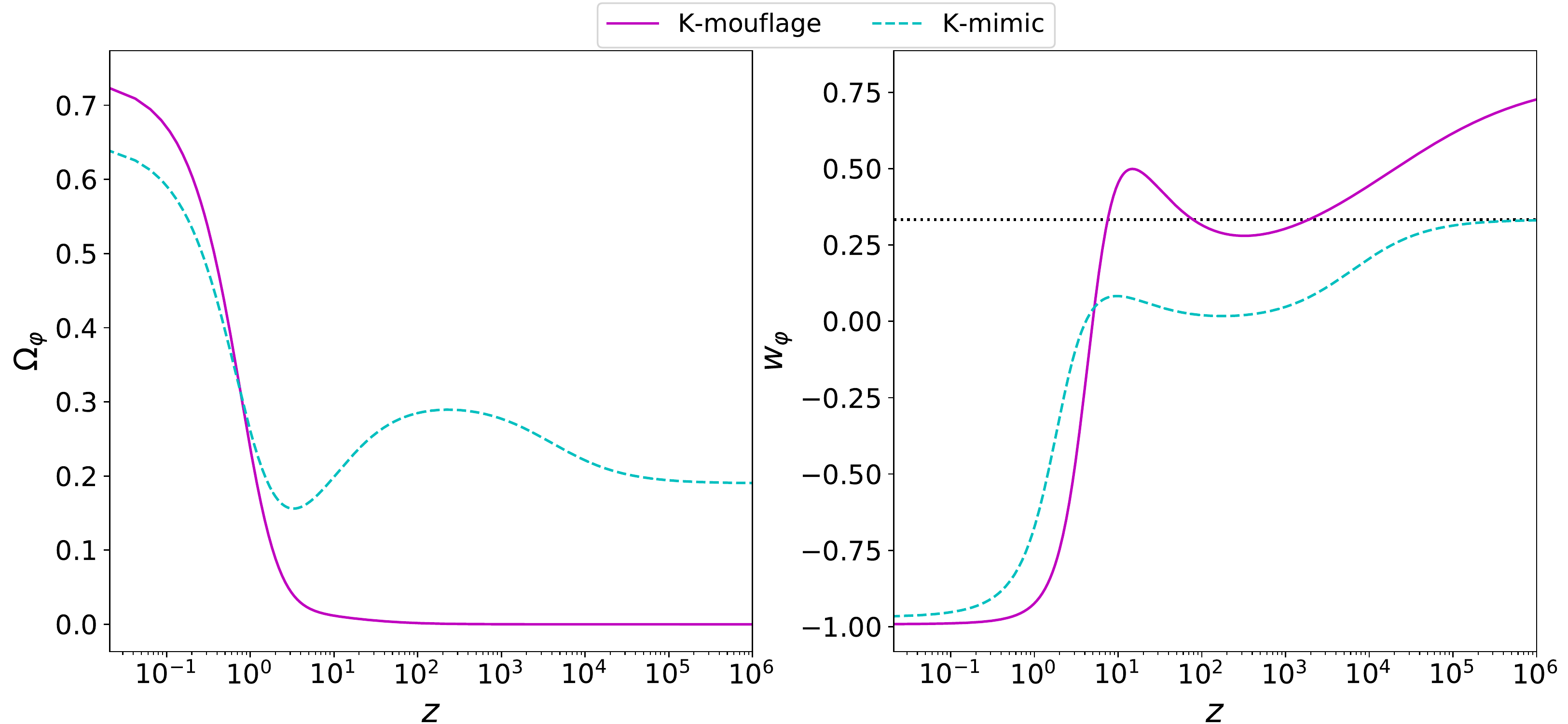}
\caption{ \emph{Left panel}: Scalar field density parameter $\Omega_{\varphi}=\frac{\rho_{\varphi}}{3 M_{pl} H^2}$. \emph{Right panel}: Scalar field equation of state $w_{\varphi}=\frac{p_{\varphi}}{\rho_{\varphi}}$. The horizontal dotted line denotes the equation of state of radiation $w=1/3$.  We consider a K-mouflage model with parameters \{$\alpha_U=1$, $\gamma_U=1$, $m=3$, $\gamma_A=0.2$, $\epsilon_{2,0}=-2\times 10^{-2}$\} and a K-mimic model with parameters  \{$m=3$, $\gamma_A=0.2$, $\epsilon_{2,0}=2\times 10^{-2}$\}.  
}
\label{fig:omega}
\end{center}
\end{figure}  \\ The non-minimal coupling of the scalar field to matter fields, determines a running of the effective Planck mass, or equivalently of the effective Newtonian constant, which is displayed in the right panel of Fig.~\ref{fig:adotoa}. We can see that in the case of K-mouflage the effective Newtonian constant is higher than the GR value at all redshifts. For K-mimic scenarios,  in which $\bar{A}^2 <1$ and $\epsilon_1>0$, the effective Newtonian constant function is lower than in GR until very low redshifts.  
\\ Fig.~\ref{fig:omega} shows the background energy density of the scalar field in units of the critical density and its equation of state, for the same models considered in Fig.~\ref{fig:adotoa}.  As we can see, in K-mouflage models the scalar field energy density becomes completely sub-dominant for $z\gtrsim 1$  and the early-time deviation of the Hubble rate from $\Lambda$CDM is only determined by the early-time behaviour of the coupling function $\bar{A}$, being different from unity at high redshift, as required by construction of the theory \cite{2016JCAP...01..020B}. In K-mimic models, instead, the scalar field gives a non-negligible contribution to the energy content of the universe at all times. In such models, indeed, $\Omega_{\varphi}$ has to compensate for the pre-factor $(\bar{A}/(1-\epsilon_2))^2$ in the Friedmann equation~(\ref{E00-1}), which is lower than one at $z\gtrsim1$. Therefore, during the matter-dominated epoch, the scalar field behaves like pressure-less matter, while it becomes relativistic at $a<a_{eq}$, adding a further contribution to radiation. \\ In Fig.~\ref{fig:Psi_Phi} we compare the effect of K-mouflage and K-mimic gravity on the two Bardeen potentials. We see that, as expected from our discussion in Sec~\ref{sec:km}, K-mouflage models induce a late-time anisotropic stress so that $\Phi$ is enhanced w.r.t. standard GR while $\Psi$ is suppressed. In K-mimic the gravitational slip is almost absent because the factor $\epsilon_1$ is much lower than in K-mouflage, and the two Bardeen potentials  are both strongly suppressed on small and intermediate scales. Depending on the scale, the suppression can take place also at high redshift, deep in matter domination. 
\begin{figure}[tbp]
\begin{center}
\includegraphics[width=\linewidth]{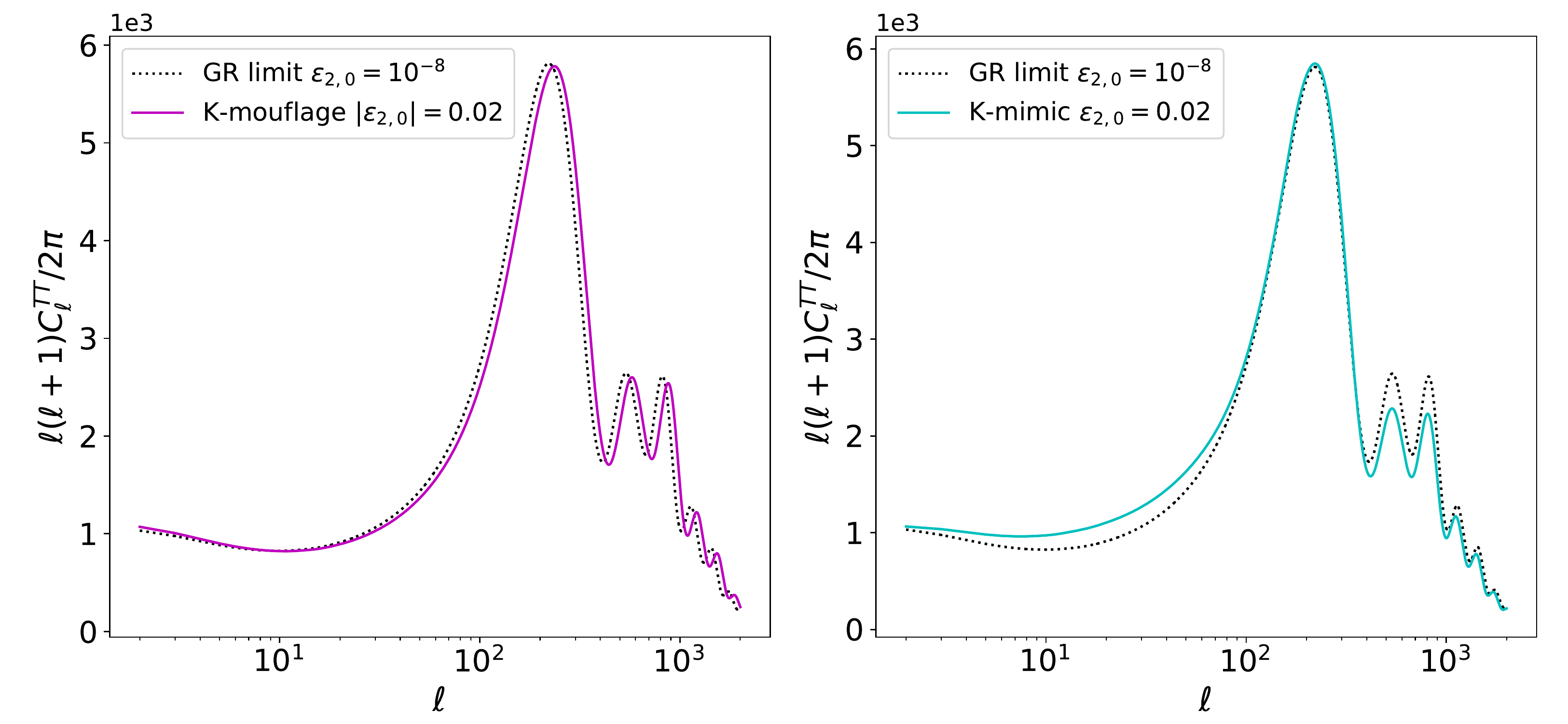}
\caption{Temperature power spectrum for K-mouflage (left panel, violet curve) and K-mimic (right panel, cyan curve) compared to the solution obtained in the $\Lambda$CDM limit $\epsilon_{2,0} \rightarrow 0$ (black curve).  We consider a K-mouflage model with parameters \{$\alpha_U=1 $, $\gamma_U=1$, $m=3$, $\gamma_A=0.2$, $\epsilon_{2,0}=-2\times 10^{-2}$\} and a K-mimic model with parameters  \{$m=3$, $\gamma_A=0.2$, $\epsilon_{2,0}=2\times 10^{-2}$\}.
}
\label{fig:TT}
\end{center}
\end{figure} \\  The effect of K-mouflage and K-mimic features on the CMB temperature power spectrum is shown in Fig.~\ref{fig:TT}. 
In K-mouflage models, acoustic peaks are shifting on the $\ell$-axis as the parameters of the models are varied. The more the parameters deviate from the $\Lambda$CDM limit, the more the peaks result shifted toward higher multipoles. This $\ell$-axis displacement is due to the change in the background expansion history, which modifies both the sound horizon scale at last scattering ($r_s$) and the comoving distance to last scattering $\tau_0 - \tau_{\star}$, where $\tau$ is the conformal time. The angular position of the peaks is with good approximation proportional to the ratio: $\frac{\tau_0 - \tau_{\star}}{r_s}$, and in K-mouflage this ratio results to be higher than in $\Lambda$CDM, determining the shift.
 In K-mimic models, the Hubble factor is modified during the matter dominated epoch, as shown in Fig.~\ref{fig:adotoa}, but the ratio $\frac{\tau_0 - \tau_{\star}}{r_s}$ remains almost constant as the parameters move away from the $\Lambda$CDM limit and we do not observe any shift in the angular position of acoustic peaks. On the other hand in the case of K-mimic, the scalar field represents a non-negligible energy source in the Einstein-Boltzmann equations. This determines a scale-dependent change in the amplitude of the CMB power spectrum. At low-$\ell$, before the first acoustic peak the power spectrum is boosted by an enhanced early-ISW effect, that is determined by the strong suppression of the Weyl potential in deep matter-domination. At high-$\ell$, beyond the first peak, we observe a decrease of power related to the decreased gravitational force that drives acoustic oscillations in the tight-coupled regime. \\ 
In Fig.~\ref{fig:rel_dev} we show the deviation of K-mouflage and K-mimic power spectra from the $\Lambda$CDM solution. We consider three cosmological probes: the CMB temperature, the CMB lensing potential and the dark-matter density fluctuations. Before analysing in detail the effect of the different parameters, we discuss the general effect of K-mouflage and K-mimic on the matter power spectrum and on the lensing potential power spectrum.
\\The $P(k)$ behaviour in the two scenarios is linked to the different evolution of the Hubble rate and of the Newtonian Constant $G_{N,eff}$, displayed in Fig.~\ref{fig:adotoa}. Computing the $P(k,z)$ at $z>2$ (i.e. in the matter dominated epoch, where the dynamics of the scalar field negligible and $\epsilon_1 \sim \epsilon_2 \sim 0$) we verified that on large scales, above the Hubble radius, both K-mimic and K-mouflage show a negative deviation of $P(k,z)$ w.r.t. $\Lambda$CDM, which is of order 10\% for $\epsilon_{2,0} \sim 10^{-2}$. This behaviour is directly related to the background expansion history, that affects the dynamics of perturbations on super-horizons scales. The positive deviation of the Hubble rate w.r.t. $\Lambda$CDM, displayed in Fig.~\ref{fig:adotoa} leads to a damping of super-horizon perturbations in both K-mouflage and K-mimic, that manifests with a reduced $P(k,z)$ at small $k$ and high $z$. 
 \\ This effect can also be understood in terms of a change in the initial conditions for matter perturbations. The adiabatic growing mode in synchronous gauge, that is used in EFTCAMB as initial condition, is given by Eq.~(22) of \citep{Bucher}. The initial perturbation in the dark matter fluid in synchronous gauge is proportional to the square conformal time $\delta \sim \tau^2$. Computing the relative deviation in $\tau^2$ w.r.t. to the $\Lambda$CDM solution for K-mimic and K-mouflage, one recovers a negative deviation of the same order for both models. \\
After the modes have entered the horizon, they feel the effect of the enhanced $G_{N,eff}$ for K-mouflage, this leads to an enhanced clustering, so that the $P(k)$  rises, at z=0 all scales of interest have entered the horizon so that we see a positive $\Delta P/P$ almost everywhere, depending on the choice of the parameters, especially the $\alpha_U$ parameter, as we are going to discuss. In K-mimic, the modes inside the horizon feel the effect of a reduced $G_{N,eff}$, that damps the growth of perturbations compared to the $\Lambda$CDM case. The deviation from  $\Lambda$CDM in both models is larger for high-k modes that have entered the horizon when $G_{N,eff}$ was farther away from the $\Lambda$CDM limit than it is at $z \simeq 0$. \\
On the other hand the CMB lensing probes the clustering at redshift up to $10$, so for K-mouflage, the lensing potential power spectrum keeps track of the negative $\Delta P /P$ at large scales (low multipoles), while it shows an increase on large multipoles. In K-mimic we observe a negative deviation at all multipoles, corresponding to the negative $\Delta P /P$. \\
To interpret the effect of the different parameters defined in Sec~(\ref{sec:parameters}) on the cosmological observables, we compare the predictions of different models in terms of relative difference w.r.t. the $\Lambda$CDM limit. 
Since all models with $\epsilon_{2,0} \to 0$ converge to the standard $\Lambda$CDM cosmology, we investigate the impact of modifying other parameters by fixing $\epsilon_{2,0} = 10^{-2}$, a value consistent with Solar System constraints, and varying them one by one. \\
 Taking the red line as reference, we see that varying the value of the $m$ parameter (blue curve), has close to no impact on the different spectra, for both K-mouflage and K-mimic therefore we expect this parameter to be almost unconstrained from data. \\ Increasing the value of $\gamma_A$ (green curve), seems to push the spectra toward the $\Lambda$CDM limit. Indeed, taking the limit $\gamma_A \rightarrow \infty$ in the definition of $\bar{A}(a)$ Eq.~(\ref{A_def}) gives $\bar{A}\rightarrow1 -(1-a^{\nu_A}) \epsilon_{2,0}/\nu_A $, which remains close to $1$ for typical values of $\epsilon_{2,0}$ (the exponent $\nu_A$ can vary between 1 and 1.5). We thus expect data to show some degree of degeneracy between $\epsilon_{2,0}$ and $\gamma_A$. \\
The parameters $\alpha_U$ and $\gamma_U$ that control the late-time behaviour of the kinetic function in K-mouflage models have small impact on the cosmological observables. The spectra showed in Fig.~\ref{fig:rel_dev} are almost totally insensitive to the $\gamma_U$ parameter, which can then be safely fixed in future analysis. The parameter $\alpha_U$ affects the evolution of large scales perturbations, as it is only important at late times. An increasing value of $\alpha_U$ pushes the kinetic function toward the cosmological constant behaviour at higher and higher redshift, leading to a larger suppression of the gravitational potential on large scales and to an enhanced late-ISW effect. \\ As a concluding remark, we note that early-time probes, like the CMB temperature power spectrum, are more suitable for constraining K-mouflage models than K-mimic, due to the early modification of the background expansion. This determines the horizontal shift in the acoustic peaks, which is the dominant observable effect. On the other hand K-mimic models display strong growth suppression of perturbations during the matter-dominated epoch, heavily affecting late-time observables, such as matter power spectrum and CMB lensing.

\section{Parameter constraints} \label{sec:constr}
The parameters of the K-mouflage model can be constrained by current and future CMB and large scale structure data. 
In this section we present the formalism for constraining the parameters of the model by performing Fisher Matrix forecasts, as well as a full MCMC analysis using EFTCosmoMC \cite{Raveri2014}.  
\subsection{Fisher Matrix Forecasts}\label{sec:fisher}
In the following paragraphs we give a brief description of our Fisher Matrix forecasts for K-mouflage parameters with future CMB surveys. We consider a parameter space consisting of the standard $\Lambda$CDM parameters together with the K-mouflage parameters, 
\begin{equation}
\label{eq:par}
\textbf{P}=\{ \Omega_b h^2, \Omega_c h^2, H_0, n_s, \tau, A_s\} \cup \{\alpha_U, \gamma_U, m, \epsilon_{2,0}, \gamma_A\} \, .
\end{equation}
We determine the CMB power spectrum in multipole space ($C_l$'s) in the K-mouflage model with the extension to the EFTCAMB code discussed in Appendix \ref{Appendix}. We consider the following temperature and polarisation channels for the power spectra: $TT$, $EE$, $TE$, $dd$, $dT$ and $dT$, where $T$ is the temperature, $E$ -- the $E$-mode polarisation and $d$ -- the deflection angle. 

Assuming Gaussian perturbations and Gaussian noise, the Fisher matrix is then calculated as
\begin{equation}
\label{eq:fisher}
F_{ij}=\sum_l \sum_{X,Y} \frac{\partial C_l^X}{\partial p_i} (\text{Cov}_l)_{XY}^{-1} \frac{\partial C_l^Y}{\partial p_j} \, ,
\end{equation}
where the indices $i$ and $j$ span the parameter space \textbf{P} from Eq.~(\ref{eq:par}), $X$ and $Y$ represent the channels considered and $\text{Cov}_l$ is the covariance matrix for multipole $l$. In calculating the covariance matrix, the instrumental noise must be considered. Given the instrumental noise for the temperature and $E$-polarisation channel, the noise corresponding to the deflection angle can be determined through lensing reconstruction using the minimum variance estimator \cite{2002ApJ...574..566H}. 
The covariance matrix is discussed in detail in Ref. \cite{1999ApJ...518....2E}, where its elements are given explicitly [Eqs. (4)-(11)]. 

We consider the \textit{Planck} 2015 \cite{2016A&A...594A..13P} values as the fiducial values to the $\Lambda$CDM parameters, while for K-mouflage we test a few scenarios.  

We consider two space probes, \textit{Planck} \cite{2016A&A...594A...1P} and \textit{CORE} \cite{2018JCAP...04..016F}. We anticipate that the K-mouflage models can be tightly constrained with existing CMB data from \textit{Planck}, as actual data analysis will confirm in the next section. We then show that the constraints can be significantly improved in the future with \textit{CORE}, by around one order of magnitude. Noise specifications for \textit{CORE} can be found in \cite{2018JCAP...04..016F}.

When considering a fiducial value of $\epsilon_{2,0}=-10^{-8}$, the other four K-mouflage parameters are almost unconstrained, and in the \textit{Planck} scenario the $\sigma_{(\epsilon_{2,0})} \sim 10^{-3}$. Full forecasts for the two probes are presented in Table \ref{tab:fc1}.

\begin{table}
\begin{center}
\caption{Forecasts for a $\Lambda$CDM-like K-mouflage model, with fiducial value $\epsilon_{2,0}=-10^{-8}$, together with $\Lambda$CDM constraints.
}
\medskip
\begin{tabular}{   c c c c c c}
\hline
\multicolumn{6}{ c }{CMB experimental specifications}\\
 Parameter    &  Fiducial value &  $\sigma_{\text{Planck}}$ & $\sigma_{\text{Planck}}$ &  $\sigma_{\text{CORE}}$ & $\sigma_{\text{CORE}}$ \\
 $\alpha_U$      &  0.1         &  598  & --  &13 & --  \\
 $\gamma_U$      &  1           & 2789  & --  &65 & --\\
 $m$             &  3           &  5411 & --  & 207& -- \\
 $\epsilon_{2,0}$    &  $-10^{-8}$  &  $1.69 \times 10^{-3}$ & -- & $1.01 \times 10^{-4}$ & -- \\
 $\gamma_A$      &  0.2         &  39.30 & -- & 16.57 & -- \\
 \hline
 $\Omega_bh^2$   & 0.02226      & $2.12 \times 10^{-4}$  & $1.79 \times 10^{-4}$  & $2.58 \times 10^{-5}$  & $2.45 \times 10^{-5}$\\
 $\Omega_ch^2$   & 0.1193       & $1.48 \times 10^{-3}$  & $1.44 \times 10^{-3}$  & $4.99 \times 10^{-4}$  & $4.82 \times 10^{-4}$ \\
 $H_0$           & 67.51        & 2.51                   & 0.76                    & 0.23 &  0.21 \\
 $n_s$           & 0.9653       & $5.90 \times 10^{-3}$  & $4.42 \times 10^{-3}$   &  $1.41 \times 10^{-3}$ & $1.41 \times 10^{-3}$\\
 $\tau$          & 0.063        & $4.23 \times 10^{-3}$  & $4.25 \times 10^{-3}$ &  $1.91 \times 10^{-3}$ & $1.94 \times 10^{-3}$\\
 $A_s$           & $2.1306 \times 10^{-9}$ & $1.83 \times 10^{-11}$  & $1.79 \times 10^{-11}$ & $8.30 \times 10^{-12}$ & $8.27 \times 10^{-12}$ \\
\hline
 \end{tabular}
\label{tab:fc1}
\end{center}
\end{table}

\subsection{Markov-Chain-Monte-Carlo constraints}

\begin{table}[h]
\begin{center}
\begin{tabular}{l | cccc}
parameter & CMB & CMB+CMBL & CMB+CMBL+SN+BAO & ALL \\
\hline
$|\epsilon_{2,0}|$ & $<0.04$ & $<0.04$ & $<0.04$ & $<0.042$ \\
$\gamma_A$ & $-$ & $-$ & $-$ & $-$ \\
$\alpha_U$ & $0.4^{+1.0}_{-0.42}$ & $0.4^{+1.0}_{-0.42}$ & $0.31^{+0.59}_{-0.31}$ & $0.41^{+0.91}_{-0.41}$ \\
$\gamma_U$ & $-$ & $-$ & $-$ & $-$\\
$m$ & $-$ & $-$ & $-$ & $-$  \\
\hline
$H_0$ & $70.1^{+4.1}_{-3.4}$ & $70.3^{+4.1}_{-3.4}$ & $70.1^{+3.2}_{-2.6}$ & $71.5^{+3.3}_{-3.1}$ \\
$\sigma_8\Omega_m^{0.5}$ & $0.46^{+0.02}_{-0.02}$ & $0.45^{+0.016}_{-0.015}$ & $0.45^{+0.013}_{-0.012}$ & $0.45^{+0.012}_{-0.012}$ 
\end{tabular}
\caption{
\label{table:parameters_kmouflage}
The 95\% C.L. marginalized constraints on the K-mouflage model parameters, the Hubble constant $H_0$ and $\sigma_8\Omega_{\rm m}^{0.5}$.
We do not report the constraints on parameters that are compatible with the prior at 95\% C.L..
}
\end{center}
\end{table} 

\begin{table}[h]
\begin{center}
\begin{tabular}{l | cccc}
parameter & CMB & CMB+CMBL & CMB+CMBL+SN+BAO \\
\hline
$\epsilon_{2,0}$ & $<2.1 \cdot 10^{-3}$ & $<2.4  \cdot 10^{-3}$ & $<2.3  \cdot 10^{-3}$ \\
$\gamma_A$ & $-$ & $-$ & $-$ \\
$m$ & $1.6^{+1.9}_{-0.61}$ & $1.4^{+1.1}_{-0.44}$ & $1.5^{+1.3}_{-0.53}$ \\
\hline
$H_0$ & $67.4^{+1.4}_{-1.3}$ & $67.5^{+1.2}_{-1.3}$ & $67.9^{+0.9}_{-0.9}$  \\
$\sigma_8\Omega_m^{0.5}$ & $0.46^{+0.02}_{-0.02}$ & $0.45^{+0.016}_{-0.015}$ & $0.45^{+0.014}_{-0.013}$ 
\end{tabular}
\caption{
\label{table:parameters_K-mimic}
The 95\% C.L. marginalized constraints on the K-mimic model parameters, the Hubble constant $H_0$ and $\sigma_8\Omega_{\rm m}^{0.5}$.
We do not report the constraints on parameters that are compatible with the prior at 95\% C.L..
}
\end{center}
\end{table}

\begin{figure}[tbp]
\begin{center}
\includegraphics[width=\columnwidth]{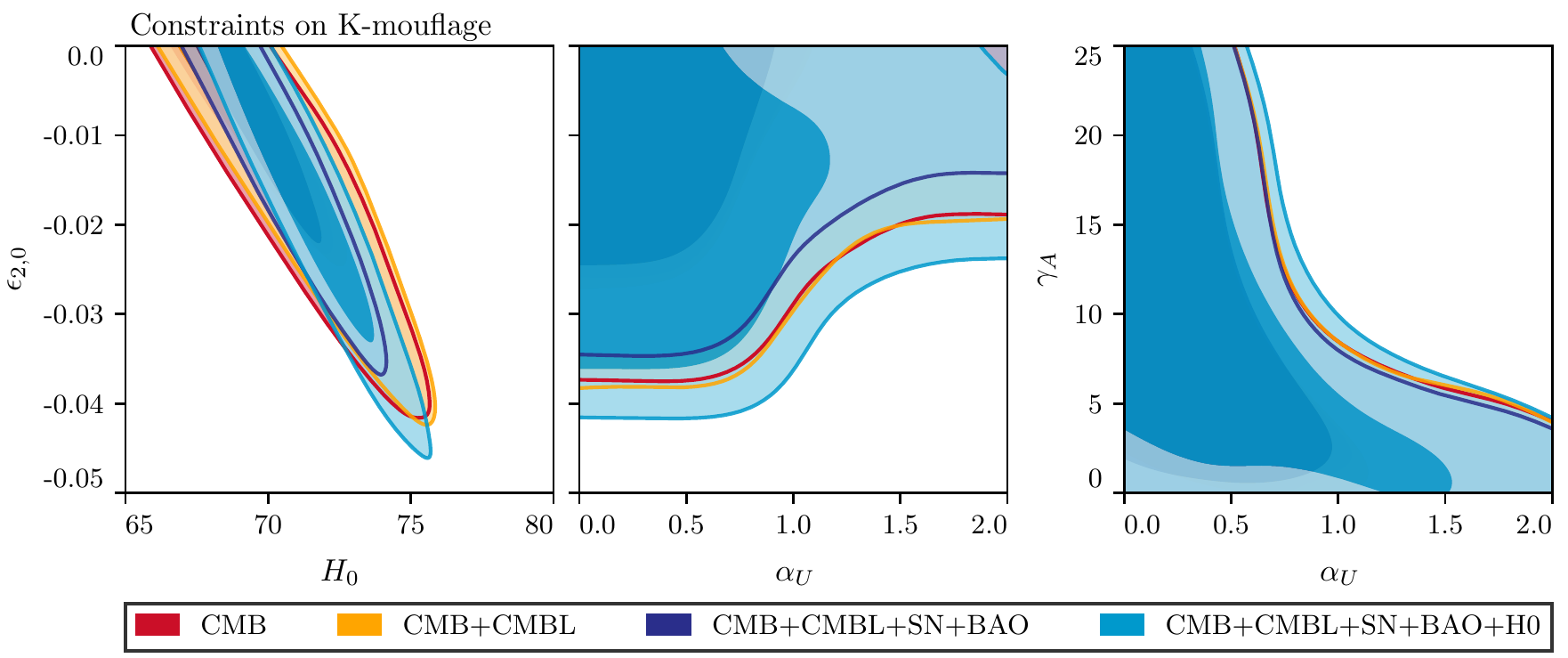} \\
\caption{
The marginalized joint posterior for a subset of parameters of the K-mouflage model and the Hubble constant.
In all three panels different colors correspond to different combination of cosmological probes, as shown in legend. 
The darker and lighter shades correspond respectively to the 68\% C.L. and the 95\% C.L. regions. 
}
\label{fig:epstest_Km}
\end{center}
\end{figure}

To constrain K-mouflage parameters from actual data, we use {\it Planck} measurements of CMB fluctuations in temperature (T) and polarization (E,B)~\cite{2016A&A...594A..13P, Aghanim:2015xee}, denoting this data set as the {\it CMB} one. In addition, we consider the {\it Planck} 2015 full-sky measurements of the lensing potential power spectrum~\cite{Ade:2015zua} in the multipoles range $40\leq \ell \leq 400$ and denote this data set as the {\it CMBL} one. We exclude multipoles above $\ell=400$ from the analysis, as CMB lensing, at smaller angular scales, is strongly influenced by the non-linear evolution of dark matter perturbations. 
We further include the ``Joint Light-curve Analysis'' (JLA) Supernovae sample~\cite{Betoule:2014frx}, which combines SNLS, SDSS and HST supernovae with several low redshift ones and BAO measurements of: BOSS in its DR12 data release~\cite{Alam:2016hwk}; the SDSS Main Galaxy Sample~\cite{Ross:2014qpa}; and the 6dFGS survey~\cite{Beutler:2011hx}.
These data sets allow breaking geometric degeneracies between cosmological parameters as constrained by CMB measurements.
All the previous data sets are complemented by the 2.4$\%$ estimate of the Hubble constant ($H_0$) by \cite{Riess_2016}.
We join all these data sets together in a data set that we denote as {\it ALL}.

We sample the parameter posterior via Monte Carlo Markov Chain (MCMC), using CosmoMC~\cite{Lewis:2002ah} in its modified version EFTCosmoMC~\cite{Raveri2014}.

Marginalized bounds on model parameters are summarized for all cases in Tables~\ref{table:parameters_kmouflage} and \ref{table:parameters_K-mimic} for K-mouflage and K-mimic models respectively.

From Table~\ref{table:parameters_kmouflage} one can notice that the constraints on the $\epsilon_{2,0}$ parameter for K-mouflage, are comparable with those derived by Solar System tests. In particular $\vert \epsilon_{2,0} \vert$ is constrained to be smaller than $0.04$, at $95\%$ C.L., from CMB data only, and the addition of CMBL, SN and BAO does not lower this bound sensibly, showing that the most of the constraining power comes from early time probes, as expected.
Remarkably, when we add local measurements of $H_0$, the constraint on $\epsilon_{2,0}$  become looser, showing that there is a degeneracy between these two parameters. 
This degeneracy is evident from the first panel of Fig.~\ref{fig:epstest_Km}, where we see the marginalized joint posterior of $\epsilon_{2,0}$ and $H_0$. At the leading order, a decrease of $\epsilon_{2,0}$, which is negative in K-mouflage, can be balanced by an increase of $H_0$, since the two parameters shift the acoustic peaks of the CMB power spectrum in opposite directions. 
This means K-mouflage models can mitigate the tension between CMB estimates and direct measurements of $H_0$ via distance ladder, that is found at about $3 \sigma$ in $\Lambda$CDM. 
Notice that the statistical significance of the tension is lowered but is not directly translated into a significant detection of $\epsilon_{2,0}$.
The K-mouflage model parameters are in fact largely degenerate and thus lower the statistical power of CMB constraints on $H_0$, as can be seen from Table \ref{table:parameters_kmouflage} and as confirmed by the MCMC.
The same argument applies to many of the K-mouflage model parameters that result in similar effects, as discussed in Sec.~\ref{sec:ps}, and are thus found to be largely unconstrained.
In particular $\gamma_A$, $m$ and $\gamma_U$ are compatible with the prior at $95\%$ C.L. Apart from $\epsilon_{2,0}$, the only K-mouflage parameter which we find to be fairly constrained by data is $\alpha_U$. This parameter only affects large scales, as we have shown in Sec~\ref{sec:ps}, thus its effect is not degenerate with that of other parameters. \\
Comparing the MCMC results with the Fisher forecast in Sec.~\ref{sec:fisher} we can see that they qualitatively confirm this picture.
The forecasted error bar on the $\epsilon_{2,0}$ parameter is stronger than the actual result because of non-Gaussianities in the posterior due to the large number of weakly constrained parameters.
\begin{figure}[tbp]
\begin{center}
\includegraphics[width=\columnwidth]{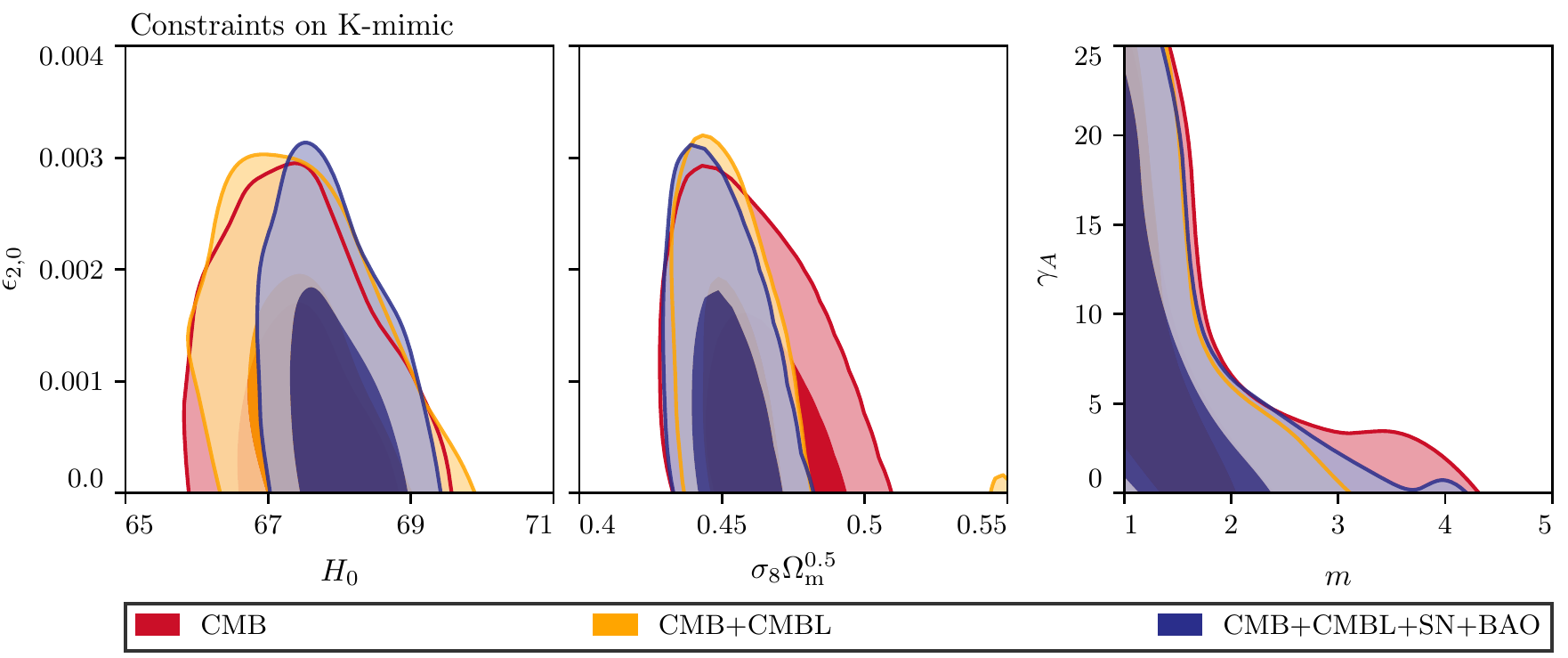} \\
\caption{
The marginalized joint posterior for a subset of parameters of the K-mimic model, the Hubble constant and $\sigma_8\Omega_{\rm m}^{0.5}$.
In all three panels different colors correspond to different combination of experiments, as shown in legend. 
The darker and lighter shades correspond respectively to the 68\% C.L. and the 95\% C.L. regions. 
}
\label{fig:epstest_Kmi}
\end{center}
\end{figure}  Furthermore these confirm that $\epsilon_{2,0}$ is the only parameter that we can significantly constrain while the other parameters of the K-mouflage model are mostly unconstrained.
The results of Table~\ref{table:parameters_kmouflage} also show that the CMB constraining power on $H_0$ is significantly lowered due to degeneracies with K-mouflage parameters.
This effect would be, however, much weaker for a \textit{CORE}-like experiment, whose observations could then be used to detect K-mouflage, at much higher statistical significance.

\begin{figure}[tbp]
\begin{center}
\includegraphics[width=\columnwidth]{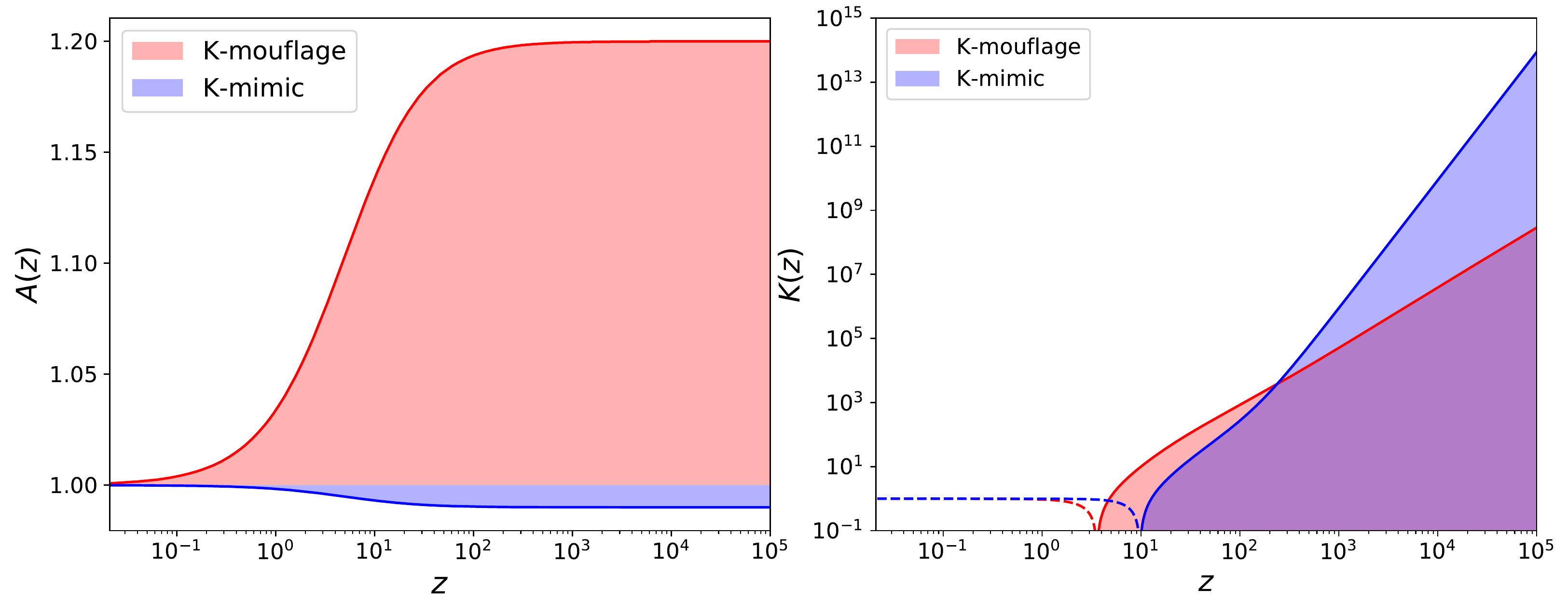} \\
\caption{
Viability regions for the coupling $A$ and the kinetic $K$ functions, expressed in terms of the redshift $z$. We consider values of the parameters at the border of the marginalized confidence intervals given in Tables~\ref{table:parameters_kmouflage} and \ref{table:parameters_K-mimic}, i.e.  \{$\alpha_U=1 $, $\gamma_U=1$, $m=2$, $\gamma_A=0.2$, $\epsilon_{2,0}=-4\times 10^{-2}$\} for K-mouflage and  \{$\alpha_U=1 $, $\gamma_U=1$, $m=2$, $\gamma_A=0.2$, $\epsilon_{2,0}= 2\times 10^{-3}$\}. The more the parameters approach the $\Lambda$CDM limit, the more the two functions move toward the constant solutions $A=1$ and $K=-1$, crossing the coloured regions.
}
\label{fig:A_K_limit}
\end{center}
\end{figure}

This picture significantly changes when we consider the K-mimic model.
As we commented in Sec~\ref{sec:mapping}, this model has an effect at the background level that can be reabsorbed by a redefinition of $\Omega_m$ but shows significant modifications of the dynamics of perturbations.
Since the constraining power of Planck measurements is higher at the level of perturbations the constraint on the $\epsilon_{2,0}$ parameter is improved as well by about one order of magnitude. Also the $m$ parameter is much more constrained, with preferred values around $2$, excluding the cubic solution $m=3$ in this scenario. 
We also notice that, since the K-mimic cosmological background is effectively unchanged, there is now no degeneracy between $\epsilon_{2,0}$ and the Hubble constant, as can be clearly seen from Fig.~\ref{fig:epstest_Kmi}.
The K-mimic model cannot be used to solve the tension between Planck measurements and distance ladder measurements.
Since the K-mimic model results in suppressed growth of late time cosmic structures, we investigate, in Fig.~\ref{fig:epstest_Kmi}, whether it is possible in this case to ease significantly the $\sigma_8$ tension. Indeed the posterior of $\epsilon_{2,0}$ and $\sigma_8\Omega_m^{0.5}$ shows a degeneracy but that is not strong enough to reconcile measurements of Planck with measurements from weak lensing surveys.  \\
The constraints shown in Tables~\ref{table:parameters_kmouflage}-\ref{table:parameters_K-mimic} can be used to infer a viability range for the coupling and the kinetic function, which is however dependent on the chosen parametrization. 
Considering extremal values for the parameters, allowed by our $95\%$ C.L. limits, we obtain a conservative estimate on how much the two functions can deviate from their $\Lambda$CDM limit according to our analysis, this is represented in Fig.~\ref{fig:A_K_limit}. We can see that the coupling function is much more constrained in K-mimic scenarios than in K-mouflage, due to the  tighter constraint on the $\epsilon_{2,0}$ parameter. In both models the kinetic function has to reproduce the cosmological constant behaviour for $z\rightarrow0$. In K-mimic the cosmological constant behaviour is reached at higher redshift than in K-mouflage, again this is a sign of the fact that the former model is more constrained by data. The large excursion of the kinetic function at very high redshift in K-mimic is related to the non-negligible scalar field energy density,  required to compensate for the pre-factor $[A/(1-\epsilon_{2,0})]^2$ in the Friedmann equation, as discussed in the previous Sections.

\section{Conclusions and Outlook}
\label{sec:concl}

In this paper we have used Cosmic Microwave Background data, in combination with BAO and SNIe, to set constraints on parameters describing K-mouflage modified gravity models.

We have employed an effective field theory description of these models and we implement two parametrisations of K-mouflage in the EFTCAMB code in order to study their effect on cosmological observables. The former is based on five parameters, where the expansion history of the Universe is free to vary, while the latter (K-mimic) has three free parameters and is forced to reproduce a close to $\Lambda$CDM background expansion. The K-mouflage and K-mimic models will be publicly released soon as part of EFTCAMB.
By varying the parameters of the models we have verified that K-mouflage can produce significant deviations in CMB angular power spectra, with respect to standard GR, and can be therefore tightly constrained by CMB probes. We have verified this via a preliminary Fisher matrix analysis, which also shows that future CMB experiments, such as \textit{CORE}, could improve K-mouflage parameter bounds currently achievable with \textit{Planck} data, by approximately one order of magnitude. For models in which the background expansion history varies, the constraining power mostly come from shifts in the position of the peaks, due to changes in the angular diameter distance to last scattering. For so called K-mimic models, in which the kinetic function of the scalar degree of freedom is chosen in such as way as to impose a degenerate expansion history with \LCDM, the most distinctive signatures come instead from variation in the linear growth rate of structures.

After this preliminary study, we have then implemented the model in the MCMC EFTCosmoMC code and derived actual parameter constraints from different data-sets, including \textit{Planck} CMB and CMB lensing, the JLA Supernovae sample and different galaxy catalogues (BOSS, SDSS and 6dFGS). The most tightly constrained parameter is $\epsilon_{2,0}$, measuring the overall departure from $\Lambda$CDM. In our analysis we have found upper limits for this parameter, which remains consistent with its $\Lambda$CDM limit ($\epsilon_{2,0}=0$) in both K-mouflage and K-mimic scenarios. These limits at 95\% C.L. are $-0.04<\epsilon_{2,0}\leq 0$ for K-mouflage and $0 \leq\epsilon_{2,0}<0.002$ for K-mimic. Some of the other model-specific parameters are unconstrained due to degeneracies with $\epsilon_{2,0}$ or due to their small impact on cosmological observables. We can however put significant bounds on $\alpha_U$, that determines the late-time behaviour of the kinetic term in K-mouflage, and on the $m$ parameter in K-mimic, that influences the behaviour of both the coupling and the kinetic term at high redshift (i.e. in the high $\tilde{\chi}$ regime). Interestingly, our analysis also shows that K-mouflage models can be used to alleviate the $H_0$ tension between \textit{Planck} and low-redshift probes, see Fig.~\ref{fig:epstest_Km} and Table~\ref{table:parameters_kmouflage}. On the other hand, K-mimic models predict a growth of matter perturbations which is slightly suppressed w.r.t. $\Lambda$CDM, resulting in lower preferred values for the $\sigma_{8}$ parameter, see Fig.~\ref{fig:epstest_Kmi} and Table~\ref{table:parameters_K-mimic}. This feature is promising to ease the tension between \textit{Planck} and weak lensing measurements, and can be further explored by running specific N-body simulations.

Neutrinos were considered to be massless in this work. In the future, we plan to generalise the study of K-mouflage both to include massive neutrinos and to assess the impact of CMB-LSS cross correlations on the constraints. In the case of the models not reproducing the $\Lambda$CDM expansion, late-time probes of the growth factor, like peculiar velocities or ISW-galaxy measurements should lead to further tightening of the constraints.

\section{Acknowledgements}
The authors are grateful to Sabino Matarrese for illuminating
discussions and comments. The authors also thank Bin Hu for useful discussions. NB, GB,  AL and ML acknowledge partial financial support by ASI Grant No. 2016-24-H.0. 
 MR is supported by U.S. Dept. of Energy contract DE-FG02-13ER41958. GB also thanks Arianna Miraval Zanon and Andrea Ravenni for comments on the draft and discussions.
\section{K-mouflage implementation in the EFTCAMB code}
\label{Appendix}

In this paper we investigate cosmological perturbations at the linear level in K-mouflage scenarios using the EFTCAMB patch of the public Einstein-Boltzmann solver CAMB. For the implementation of the model in the EFTCAMB code we adopted the so called "full-mapping" approach. In these scheme the mapping relations between the K-mouflage and the EFT action, along with the cosmological and model parameters, are fed to a
module that solves the cosmological background equations, for the specific theory, and outputs the time evolution of the EFT functions.
These functions are then used to evolve the full perturbed Einstein-Boltzmann equations and compute cosmological observables. EFTCAMB evolves the full equations for linear perturbations without relying on any quasi-static approximation. \\
For our purposes, we implemented two different versions of the model in the EFTCAMB solver, with different background evolutions, the user can switch between the two by setting the logical variable \textbf{K-mimic}. \\ If \textbf{K-mimic=F} the background expansion history is left free to deviate from the $\Lambda$CDM and the user has to fix both the $\bar{A}(a)$ and $K(a)$ functions. A model of K-mouflage is then completely specified by the choice of the standard cosmological parameters (namely $H_0$, $\Omega_{m0}$, $\Omega_{b0}$, $n_s$, $A_s$, $\tau$) and by the five additional parameters: $\alpha_U , \ \gamma_U , \ m, \ \epsilon_{2,0} , \ \gamma_A$ introduced in Section \ref{sec:parameters}. The code computes the functions $A(a)$ and $U(a)$ using the definitions in Eq.~(\ref{A_def}) and Eq.~(\ref{U_def}) and normalizing the $U(a)$ function at the present time
\begin{equation}
U(a=1)=\sqrt{\frac{-\bar{\rho}_{m0} \ \epsilon_{2,0}}{{\cal M}^4 2 \left(-3 \epsilon_{2,0}+\frac{d \ln U}{d \ln a}\vert_{a=1}\right)}} \ .
\end{equation} 
Once the function $U(a)$ and $\bar{A}(a)$ are specified, the function $\bar{\tilde\chi}$ can be computed from Eq.~(\ref{chi_tilde_U}), where the mass scale of the scalar field ${\cal M}^4$ is  fixed by the choice of the cosmological parameters
\begin{equation}
\frac{{\cal M}^4}{\bar{\rho}_{m0}}= \frac{\Omega_{\varphi 0}}{\Omega_{m 0}}+ \frac{\epsilon_{2,0}}{-3 \epsilon_{2,0} + \frac{\der \ln U}{\der \ln a}\vert_{a=1}} \ .
\end{equation} 
The code integrates the differential equation (\ref{eq:dK/dchi}) to compute $\bar{K}(a)$, with the initial condition $K_0=-1$ at $a=1$, and the background evolution is completely specified by Eq.~(\ref{E00-1}) and Eq.~(\ref{Eii-1}). \\
Otherwise if the user sets the flag \textbf{K-mimic=T }, the model reproduces a $\Lambda$CDM background expansion history. In this case the user has to specify, apart from the standard cosmological parameters, only the three parameters related to the background coupling function
$\bar{A}(a)$, i.e. \{$ \epsilon_{2,0}, \ \gamma_A, \ m $\}. Following the method developed in Sec.~\ref{sec:mapping}, the kinetic function $\bar{K}(a)$ is given by Eq.~(\ref{K_mimic}), where we fix
\begin{eqnarray}
&& \frac{{\cal M}^4}{\bar{\rho}_{m0}}= \frac{\Omega_{\varphi 0}}{\Omega_{m 0}} \ , \\ 
&& \hat{\Omega}_{\rm m0} = \frac{\Omega_{\rm m0}}{1-\epsilon_{2,0}} +\frac{2 \epsilon_{2,0} \gamma_A (1 -  \nu_{A} + 2 \Omega_{\gamma 0})+ 4 \epsilon_{2,0}(1 + \Omega_{\gamma 0})}{3(1 + \gamma_A)(1-\epsilon_{2,0})} +\frac{ \epsilon_{2,0}}{(1-\epsilon_{2,0})} \ ,
\label{Omega-m-value}
\end{eqnarray}
together with $\hat{\Omega}_{b0}=\Omega_{b0}$ and $\hat{\Omega}_{\gamma 0}=\Omega_{\gamma 0}$. The choice made in Eq.~(\ref{Omega-m-value}) satisfies the constraint given by Eq.~(\ref{Omega-constraint}) and sets the value of $\bar{\tilde\chi}$ today. The code then solves Eq.~(\ref{chi-a}) using Eqs.~(\ref{K_mimic})-(\ref{K_first_mimic}) and taking $\bar{K}'_0 =1$, $\bar{\tilde\chi}_0 = \epsilon_{2,0}/(2 \Omega_{\varphi 0})$ as initial condition at $a=1$. \\
Once the functions $\bar{K}$, $\bar{K}'$ and $\bar{\tilde\chi}$ are determined, the code solves the Friedmann equation~(\ref{E00-1}) to determine $a(t)$, using the standard $\Lambda$CDM solution as initial condition at $a \ll 1$. \\
The mapping of action (\ref{eq:actkm}) in terms of EFT functions that we reported in Sec.~\ref{sec:EFT_theory} cannot be used directly in the EFTCAMB code, that adopts a slightly different convention, according to Ref. \cite{Hu2014} . Comparing the K-mouflage action in unitary gauge and Jordan frame, Eq.~(3.11) of Ref.~\cite{2016JCAP...01..020B}, with Eq.~(1) and Eq.~(2) of Ref.~\cite{Hu2014}, we can identify the following correspondences between the EFTCAMB functions and K-mouflage

\begin{align}
\Omega(a)=& \bar{A}^{-2} -1 \ , \\
\Omega'(a)=& -2\bar{A}^{-3} \bar{A}' \ ,\\
\Omega''(a)=& 6 \bar{A}^{-4} (\bar{A}')^2 -2\bar{A}^{-3} \bar{A}'' \ , \\
\Omega'''(a)=& -24\bar{A}^{-5} (\bar{A}')^3 +18 \bar{A}^{-4} \bar{A}' \bar{A}'' -2\bar{A}^{-3} \bar{A}'''  \ ,
\end{align}
\begin{equation}
\frac{\Lambda(a) a^2}{m_{0}^{2}}= \frac{a^2 \mathcal{M}^4 \bar{K}}{m_{0}^{2} \bar{A}^4 }-\frac{3  \mathcal{H}^2 \epsilon_{2}^2}{\bar{A}^{2} }   \ ,
\end{equation}
\begin{align}
&\frac{\dot{\Lambda}(a) a^2}{m_{0}^{2}} = \frac{\mathcal{H}}{m_{0}^{2} \bar{A}^5} \left(-4 a^3 {\cal M}^4 \bar{A}' \bar{K}+a^3 {\cal M}^4 \bar{A} \bar{\tilde{\chi}}' \frac{\der \bar{K}}{\der \bar{\tilde{\chi}}} \right. \nonumber
\\ &+ \left. 6 a m_{0}^{2} \bar{A}^2 \epsilon_{2}^{2} \mathcal{H}^{2} \bar{A}'+6 m_{0}^{2} \bar{A}^3 \epsilon_{2} \mathcal{H} \left(\epsilon_{2} \left(\mathcal{H}-a \mathcal{H}'\right)-a \mathcal{H} \epsilon_{2}'\right. \right)  \ ,
\end{align}

\begin{equation}
\frac{c(a) a^2}{m_{0}^{2}}=\frac{a^2 {\cal M}^4 \bar{\tilde{\chi}}  \frac{\der \bar{K}}{\der \bar{\tilde{\chi}}}}{m_{0}^{2} \bar{A}^4}-\frac{3 \epsilon_{2}^{2} \mathcal{H}^{2}}{\bar{A}^2} \ ,
\end{equation}

\begin{align}
&\frac{\dot{c}(a) a^2}{m_{0}^{2}}= \frac{c'(a) \mathcal{H} a^3}{m_{0}^{2}} =\frac{\mathcal{H} \left(-4 a^3 {\cal M}^4 \bar{\tilde{\chi}} \bar{A}' \frac{\der \bar{K}}{\der \bar{\tilde{\chi}}} + a^3 {\cal M}^4 \bar{A}  \bar{\tilde{\chi}}' \left(\bar{\tilde{\chi}} \frac{\der^2 \bar{K}}{\der \bar{\tilde{\chi}}^2}+ \frac{\der \bar{K}}{\der \bar{\tilde{\chi}}}\right)\right)}{m_{0}^{2} \bar{A}^5} \nonumber \\ 
&+ \mathcal{H} \left(6 a \bar{A}^{-3} \epsilon_{2}^{2} \mathcal{H}^2 \bar{A}'+6 \bar{A}^{-2} \epsilon_{2} \mathcal{H} \left(\epsilon_{2} \left(\mathcal{H}-a \mathcal{H}'\right)-a \mathcal{H} \epsilon_{2}'\right)\right) \ ,
\end{align}

\begin{align}
&\gamma_1(a)= \frac{M^{4}_{2}}{m_{0}^{2} H_{0}^{2}}= \frac{\mathcal{M}^4  A^{-4}  \bar{\tilde{\chi}}^2  \frac{\der ^2 \bar{K}}{\der  \bar{\tilde{\chi}}^2}}{m_{0}^{2} H_{0}^{2}}  \ , \\
&\gamma_1'(a)= \gamma_1(a) \left(-4 \frac{\bar{A}'}{\bar{A}} + \frac{2 \chi'}{\chi} + \frac{ \chi' \frac{\der ^3 \bar{K}}{\der \bar{\tilde{\chi}}^3}}{\frac{\der ^2 \bar{K}}{\der \bar{\tilde{\chi}}^2}} \right)  \ , \\
&\frac{\der ^2 \bar{K}}{\der \bar{\tilde{\chi}}^2}=\frac{6 a^3 {\cal M}^4 \bar{\tilde{\chi}} \bar{A}' \frac{\der \bar{K}}{\der \bar{\tilde{\chi}}}-a^3 {\cal M}^4 \bar{A} \bar{\tilde{\chi}}' \frac{\der \bar{K}}{\der \bar{\tilde{\chi}}}-6 a^2 {\cal M}^4 \bar{A} \bar{\tilde{\chi}} \frac{\der \bar{K}}{\der \bar{\tilde{\chi}}}-\bar{\rho}_{m0} \bar{A}^4 \bar{A}'}{2 a^3 {\cal M}^4 \bar{A} \bar{\tilde{\chi}} \bar{\tilde{\chi}}'} \ ,
\end{align}

\begin{align}
\frac{\der ^3 \bar{K}}{\der \bar{\tilde{\chi}}^3}&=\frac{-3 (\bar{A}')^2 \bar{\tilde{\chi}}' \frac{\der \bar{K}}{\der \bar{\tilde{\chi}}}+3 \bar{A} \left(\bar{A}' (\bar{\tilde{\chi}}')^2 \frac{\der ^2 \bar{K}}{\der \bar{\tilde{\chi}}^2}+\frac{\der \bar{K}}{\der \bar{\tilde{\chi}}} \left(\bar{A}'' \bar{\tilde{\chi}}'-\bar{A}' \bar{\tilde{\chi}}''\right)\right)}{ \bar{A}^2 (\bar{\tilde{\chi}}')^2}+ \nonumber \\
&+\frac{ \frac{\der \bar{K}}{\der \bar{\tilde{\chi}}} \left(a^2 (\bar{\tilde{\chi}}')^3+6 (\bar{\tilde{\chi}}')^2 \left(a \bar{\tilde{\chi}}''+\bar{\tilde{\chi}}'\right)\right)-a \bar{\tilde{\chi}} (\bar{\tilde{\chi}}')^2 \left(a \bar{\tilde{\chi}}'+6 \bar{\tilde{\chi}}\right)\frac{\der ^2 \bar{K}}{\der \bar{\tilde{\chi}}^2} }{2 a^2 \bar{\tilde{\chi}}^2 (\bar{\tilde{\chi}}')^2} \nonumber \\ 
&-\frac{3 a \bar{\rho}_{m0} \bar{A}^2 \bar{\tilde{\chi}} (\bar{A}')^2 \bar{\tilde{\chi}}'+\bar{\rho}_{m0} \bar{A}^3 \left(a \bar{\tilde{\chi}} \bar{A}' \bar{\tilde{\chi}}''+\bar{\tilde{\chi}}' \left(\bar{A}' \left(a \bar{\tilde{\chi}}'+3 \bar{\tilde{\chi}}\right)-a \bar{\tilde{\chi}} \bar{A}''\right)\right)}{2 a^4 {\cal M}^4 \bar{\tilde{\chi}}^2 (\bar{\tilde{\chi}}')^2}  \ .
\end{align}
In the last equations we adopted the EFTCAMB notation \citep{Hu2014} where $m_{0}^{2}= \tilde{M}_{\rm Pl}^{2}$, the over-dot represents derivatives with respect to conformal time, while the prime represents derivatives with
respect to the scale factor $a$ and $\mathcal{H}(a)=a H(a)$.

\bibliography{Bib}{}

\end{document}